\documentclass[journal]{IEEEtran}
\usepackage[T1]{fontenc}
\usepackage{multirow}
\usepackage{cite}

\ifCLASSINFOpdf
   \usepackage[pdftex]{graphicx}
   \DeclareGraphicsExtensions{.pdf,.jpeg,.png}
\else
   \usepackage[dvips]{graphicx}
\fi
\usepackage{amsmath}
\usepackage[cmintegrals]{newtxmath}
\usepackage{bm}
\usepackage{array}

\ifCLASSOPTIONcompsoc
  \usepackage[caption=false,font=normalsize,labelfont=sf,textfont=sf]{subfig}
\else
  \usepackage[caption=false,font=footnotesize]{subfig}
\fi

\usepackage{url}
\usepackage[colorlinks,linkcolor=blue,citecolor=green,bookmarks=True]{hyperref}

\newcommand{\inpnglw}[4]{
	\begin{figure}[!htp]
		\setlength{\abovecaptionskip}{0cm}
		\setlength{\belowdisplayskip}{0cm}
		\centering
		\includegraphics[width=#4\textwidth]{#1}
		\caption{#2}
		\label{#3}
	\end{figure}
}

\newcommand{\inpngtlw}[4]{
	\begin{figure*}[!htp]
		\setlength{\abovecaptionskip}{0cm}
		\setlength{\belowdisplayskip}{0cm}
		\centering
		\includegraphics[width=#4\textwidth]{#1}
		\caption{#2}
		\label{#3}
	\end{figure*}
}

\usepackage{tikz}
\newcommand*\circled[1]{\tikz[baseline=(char.base)]{
		\node[shape=circle,draw,inner sep=1pt] (char) {#1};}}
\usepackage{booktabs}
\usepackage{xcolor}

\definecolor{lime}{HTML}{A6CE39}
\DeclareRobustCommand{\orcidicon}{%
	\begin{tikzpicture}
	\draw[lime, fill=lime] (0,0) 
	circle [radius=0.16] 
	node[white] {{\fontfamily{qag}\selectfont \tiny ID}};    \draw[white, fill=white] (-0.0625,0.095) 
	circle [radius=0.007];    \end{tikzpicture}
	\hspace{-2mm}}
\foreach \x in {A, ..., Z}{%
	\expandafter\xdef\csname orcid\x\endcsname{\noexpand\href{https://orcid.org/\csname orcidauthor\x\endcsname}{\noexpand\orcidicon}}
}

\usepackage{fancyhdr}
\begin{document}

\title{IRO: Integrity and Reliability Enhanced Ring ORAM}

\author{
	Wenpeng He\orcidA{}, Dan Feng, \emph{Member, IEEE}, Fang Wang, Yue Li\orcidB{}, and Mengting Lu\orcidC{}\\
	School of Computer Science and Technology, Wuhan National Laboratory for Optoelectronics \\
	Huazhong University of Science and Technology, Wuhan, 430074, China\\
	\{wenpenghe, dfeng, wangfang, yueli, mengting\_lu\}@hust.edu.cn 	
}

\maketitle

\begin{abstract}

Memory security and reliability are two of the major design concerns in cloud computing systems. State-of-the-art memory security-reliability co-designs (e.g. Synergy) have achieved a good balance on performance, confidentiality, integrity, and reliability. However, these works merely rely on encryption to ensure data confidentiality, which has been proven unable to prevent information leakage from memory access patterns. Ring ORAM is an attractive confidential protection protocol to hide memory access patterns to the untrusted storage system. Unfortunately, it does not compatible with the security-reliability co-designs. A forced combination would result in more severe performance loss. 

In this paper, we propose IRO, an Integrity and Reliability enhanced Ring ORAM design. 
To reduce the overhead of integrity verification, we propose a low overhead integrity tree RIT and use a Minimum Update Subtree Tree (MUST) to reduce metadata update overhead.
To improve memory reliability, we present Secure Replication to provide channel-level error resilience for the ORAM tree and use the mirrored channel technique to guarantee the reliability of the MUST.
Last, we use the error correction pointer (ECP) to repair permanent memory cell fault to further improve device reliability and lifetime. A compact metadata design is used to reduce the storage and consulting overhead of the ECP.

IRO provides strong security and reliability guarantees, while the resulting storage and performance overhead is very small.
Our evaluation shows that IRO only increases 7.54\% execution time on average over the Baseline under two channels four AES-GCM units setting. With enough AES-GCM units to perform concurrent MAC computing, IRO can reduce 2.14\% execution time of the Baseline. 

\end{abstract}


\thispagestyle{fancy}
\chead{This work has been submitted to the IEEE for possible publication. Copyright may be \\transferred without notice, after which this version may no longer be accessible.}
\renewcommand{\headrulewidth}{0.0pt}


\section{Introduction}

\IEEEPARstart
Memory security and reliability are two of the major design concerns of cloud computing systems. Security requires that attackers learn nothing about the user's sensitive information (confidentiality) and are not able to tamper with data content (integrity). Reliability requires that memory natural failures are detectable and correctable. 
Nowadays, many security and reliability co-design researches\cite{ivec2010,synergy2018,reducing2018,compact2020} have achieved a good balance on performance, security, and reliability. 
However, these researches merely rely on encryption to ensure data confidentiality, which has been proven unable to stop information leakage from memory access patterns. 
To overcome this privacy problem, Oblivious RAM (ORAM) is required as a general solution.
Unfortunately, the previous security-reliability co-designs are not very compatible with the ORAM.
A forced combination of the ORAM and the co-designs would result in severe performance loss and storage overhead. 
This paper investigates how to provide robust integrity and reliability guarantees for the ORAM protected memory with negligible overhead on performance and storage.

Memory encryption is wildly proposed in various secure hardware designs\cite{arch2000,sgx2016,amdsme2016} to protect data confidentiality from directly information stealing (e.g. physical memory scan\cite{memoryscan2009}, bus snooping\cite{hide2004}). But recently studies\cite{hide2004,access2012} reveal that merely encrypting data is not enough, memory access patterns can still be used to decipher sensitive information indirectly. 
For example, the memory accesses generated by two different SQLite queries running on the same database are clearly distinguishable\cite{phantom2013}. Attackers can use this information to infer content being queried. 
To completely solve this information leakage problem, ORAM\cite{theoryoram1987,softwareoram1996} is proposed. ORAM is a cryptographic primitive that can be used to hide memory access patterns by adding redundant memory transactions and random data remapping. After ORAM translation, every data request has a similar memory access pattern and is indistinguishable from the others. In this paper, we focus on the state-of-the-art ORAM design - Ring ORAM\cite{ring2015}. 

Although the ORAM provides stronger confidential protection, it was not originally designed for integrity. Without an integrity verification mechanism, the ORAM not only cannot guarantee a correct output but also leak memory access patterns\cite{pathit2013}. 
The integrity of data is usually verified by checking its associated message authentication code (MAC) that is generated by a keyed-hashing function. Any modification would cause the computed MAC does not match the stored MAC. 
However, attackers can still break through such verification by replacing current Data+MAC with an old value, which is commonly called replay attacks. Integrity tree \cite{mt2003,bmt2007,sgx2016,vault2018} is the state-of-art solution for replay attacks. It is a hash tree that constructed over data or counters. A full path in the tree is accessed to verify the integrity of a data block.

Integrate the ORAM and the integrity tree can completely protect data security in untrusted cloud environments. However, the performance degradation would be more serious. 
The external memory in Ring ORAM is structured as a binary tree. Each data request is translated to multiple reads and writes on a random path of the memory tree. 
To provide the capability to resist replay attacks, each block read in the path needs integrity verification. This causes an unacceptable integrity tree traverse overhead. 
Prior ORAM integrity verification methods\cite{pathit2013,freecursive2015,externally2017} are not suitable for the Ring ORAM in secure processor. They either unable to verify a single data block\cite{pathit2013} or unable to verify metadata/dummy blocks\cite{freecursive2015} or need large client side storage overhead\cite{externally2017}. 
How to provide an efficient and practical integrity verification scheme for the Ring ORAM in secure processors is still a problem to be solved.

Besides security protection, the reliability guarantee is also important in the memory system. Memory cell failures are very common in DRAM chips. For example, in the Jaguar system, one failure occurs approximately every six hours\cite{studydramfailures2012}.
Error Correction Code (ECC) memory is widely used in cloud servers to build a reliable memory system. It can correct single-bit errors and detect double-bit errors in 64-bit data.
However, previous studies\cite{studydramfailures2012,fengshui2013,meims2015} show that large-granularity failures are almost as common as single bit failure, and permanent memory cell damage is not a rare event. 
The ECC memory is not enough for high reliability required systems. More stronger error detection algorithms\cite{crc82005,ivec2010}, error correction methods\cite{white1997,hp2008,lotecc2012,eccparity2014,xed2016,redundant2017} and fault cell repair strategies\cite{layser2003,ecp2010,archshield2013,flower2020} are needed in commercial memory systems.

In mainframe servers, memory failures must be detected and corrected even if there is a complete failure of a DIMM\cite{ibm2012}.
The main obstacle to provide channel-level error resilience capability for the ORAM protected memory is capacity overhead. 
RAID-based techniques such as RAIM\cite{ibm2012} and Mirrored Channel\cite{hp2008} can tolerate one channel failures but results in 20\% and 50\% memory waste respectively. As the ORAM has resulted in low memory utilization (50\% in most ORAM schemes\cite{path2013,tiny2015,ring2015}), these RAID-based schemes are cost-prohibitive to deploy.  
Not only that, traditional permanent fault cell repair strategies are also not compatible with ORAM. The state-of-the-art solutions for repairing fault memory cells are using the spares to replace fault cells\cite{ecp2010}, words\cite{archshield2013} and rows\cite{layser2003}. These schemes need additional space to store the fault remapping information, and each access requires consulting the remapping table to judge and correct the failed cells. The hot cache is an efficient approach to reduce the consulting overhead\cite{archshield2013}.
But, after ORAM amplifying and randomizing memory operations, the cache becomes inefficient. Multiple fault map entries need to be accessed in each ORAM request.

To address the problems described above, this paper proposes IRO, an \textbf{I}ntegrity and \textbf{R}eliability enhanced Ring \textbf{O}RAM scheme. 
The primary contributions are listed as follows:

\begin{itemize}
	\item Similar to previous security-reliability co-designs, ECC chips are reused for secure metadata storage. The MAC in the secure metadata can be used to provide stronger error detection capability than ECC. Such a re-purposing reduces extra metadata access overhead.

	\item We propose RIT, a low overhead Ring ORAM integrity tree to defend against replay attacks. The RIT smartly combines two kinds of integrity tree (i.e. BMT\cite{bmt2007} and PIT\cite{pathit2013}). It can hide the traverse of the integrity tree in normal Ring ORAM memory accesses, and therefore reducing the integrity verification overhead.
	
	\item We propose Minimum Update Subtree Tree (MUST) to reduce metadata update overhead. After a bucket is accessed, the valid bits and read counter in its metadata need to be updated. This leads to great unnecessary writes. Thus, we separate these data from metadata blocks and groups as a subtree organization to reduce the number of blocks that need to be updated after each path access. 
	
	\item  We propose Secure Replication to achieve as strong reliability as the Mirrored Channel with negligible storage overhead. We observe that the dummy slots in Ring ORAM are wasted that can be utilized to store the replica of real blocks. We store the replica in the different channels with its original data. Thus, any failures within a channel can be corrected with other channels. For the MUST with a smaller capacity overhead, we use the Mirrored Channel techniques to guarantee its reliability.
	
	\item We use Error Correction Pointers (ECPs) to repair the failed memory cells that repeatedly return wrong data. To address the storage and consulting overhead of ECPs, we propose a compact metadata design to integrate ECPs into ORAM structures and eliminate their access. In the Ring ORAM tree, we integrate ECPs into the metadata block and use them to repair the permanent fault in the bucket. For the MUST, we have some ECPs at each MUST node to repair the permanent fault that occurs in the node and its mirror.

\end{itemize}

IRO provides stronger integrity and reliability protection for Ring ORAM with a small overhead on the memory storage and the system performance. The results show that our scheme only increases 7.54\% execution time of Ring ORAM under the two channels four AES-GCM units setting. Given enough MAC computing resources can reduce 2.14\% execution time of Ring ORAM. 

\section{Background}\label{sec_background}

\subsection{Threat Model \& Fault Model}

\textbf{Threat Model.} We assume a threat model similar to prior works\cite{path2013,ring2015,synergy2018}. The secure processor is the only trusted hardware free from physical attacks. All data inside the processor cannot be observed or tampered with by outer attackers. On the other hand, all components outside the processor (e.g. processor-memory bus and external memory) are insecure. The content, address, and operation type of memory requests are under the surveillance of attackers and can be modified unauthorizedly.
In such settings, ensure memory security needs the following techniques. 
\emph{Memory encryption} is a basic confidential protection method to avoid the memory content being illegal read. 
But, merely encryption is not enough, memory access patterns can still leak considerable sensitive information.
\emph{Oblivious RAM (ORAM)} is necessary to prevent such information leakage.  
\emph{Message authentic codes (MACs)} is common used to protect data integrity from memory content tampering. 
However, it cannot resist replay attacks that replace the Data+MAC with an old value.
To prevent such attacks, \emph{integrity tree} is required to check memory freshness.

\textbf{Fault Model.} Memory failures are very common in modern compute clusters and can be caused by transient faults and permanent faults.
\emph{Transient faults}, which randomly corrupts data and not result in hardware damage. The transient faults can be corrected by overwritten correct data. The errors caused by such faults are usually called soft errors. Data tampering can also be treated as a type of soft errors. 
\emph{Permanent faults}, also called hard faults that are usually caused by memory cell damage and cannot be repaired at the hardware. We call the repeated errors caused by the hard faults hard errors.
Similar to the threat model, we assume the processor is error-free, and the external memory needs error detection and correction methods.
Nowadays, most commercial memory systems are at least using SECDED ECC memory to protect data correctness from one-bit error in 64-bit data. For systems that require high reliability, stronger error correction methods (e.g. Mirrored Channel) and memory repair strategies (e.g. Error Correction Pointer) are required and adopted.

\subsection{Confidentiality Protection}
\textbf{AES counter mode encryption (AES-CTR)}\cite{aesctr2000} is a state-of-the-art memory encryption method for secure memory systems.
In the AES-CTR, each memory block is associated with an encryption counter. The block address and the counter is encrypted using AES to generate a One Time Pad (OTP). The OTP is XORed with the plaintext/ciphertext to generate ciphertext/plaintext. To avoid reuse of the OTP, the counter is incremented on each data re-encryption.

\textbf{Ring ORAM}\cite{ring2015} is a most efficient ORAM protocol for the secure processor to obfuscate memory access patterns. Its hardware components are constructed by an untrusted encrypted memory and a trusted on-chip ORAM controller.
The external encrypted memory is structured as a binary tree called ORAM tree. Each node (called bucket) in the tree contains some security metadata (i.e. \emph{addresses, path labels, offsets, valid bits, read counter and encryption counter}), \emph{Z} real slots for storing real block and \emph{S} dummy slots for storing dummy blocks. The location of the slots are randomly permuted to make the encrypted real and dummy blocks indistinguishable. The location of real blocks is identified with the \emph{offset}. The \emph{valid bits} (VBits) are used to mark which blocks have been accessed, and the \emph{read counter} (ReadCtr) records how many times the bucket has been accessed since it was last written. The \emph{encryption counter} (EncCtr) is used to encrypt blocks with AES-CTR.
The ORAM controller includes two main components: the position map (PosMap) and stash. The PosMap is a look-up table that records which path in the ORAM tree each real block located. Ring ORAM guarantees an invariant: a block labeled with \emph{path-l} in the PosMap must reside somewhere on the \emph{path-l} or in the stash. The stash is a small on-chip buffer that stores data blocks temporarily.

The Ring ORAM protocol contains three main operations.

\emph{Read Path}: When the on-chip cache/buffer miss occurs, the target block is read from memory follows the following steps. \circled{1} The path label \emph{l} of the target is looked up in the PosMap and updated with a new random path label. \circled{2} The metadata block in each bucket on \emph{path-l} is read first to find out which bucket contains the target. \circled{3} One valid block (i.e. the block not read since last written) either the target or a random dummy block is read from each bucket on the path. \circled{4} Update the VBits and ReadCtr in the accessed metadata blocks in the memory.

\emph{Evict Path}: To reduce stash occupancy, after every \emph{A} times of \emph{Read Path}, a random path in the ORAM tree is selected to store the evicted stash blocks. \emph{Z} valid blocks in each bucket on the path are read to the stash. Then, the blocks in the stash are greedily evicted to the path after shuffling and encryption. 

\emph{Early Reshuffle}: To ensure randomness and security,  once a bucket is read \emph{S} times since it was last written, the early reshuffle is performed. It reads \emph{Z} valid blocks and writes \emph{Z+S} shuffled blocks to the bucket just like in the \emph{Evict Path}.

\subsection{Integrity Verification}
\textbf{Message authentication code (MAC)} \cite{hmac1996} is created using a keyed hashing function to provide integrity guarantees for memory blocks (i.e. MAC = Hash$_{key}$(Address, Data)). When the data and MAC are fetched to the processor, the MAC of the data is re-computed. The tampering of the address or data will cause the computed MAC to not match the one fetched from memory. To lower the MAC-collision probability, commonly acceptable lengths of MAC for the 64-byte data block are 54-bit\cite{ivec2010,morph2018},56-bit\cite{sgx2016} and 64-bit\cite{mac64b2006}. 

\textbf{Merkle Tree (MT)} \cite{mt2003} prevents replay attacks through constructing a MAC tree over data and EncCtrs (as shown in figure~\ref{it}(a)).  
Each leaf node in MT is a MAC of a data or counter block. Each MAC value in an internal node is the keyed-hash value of all its child nodes. The root of the MT is stored on the trusted computing base (TCB) and cannot be tampered with. To verify the integrity of a block, all MACs on the path where the block located are calculated. Any tampering on the path would cause a mismatch in the root MAC value. 

\textbf{Bonsai Merkle Tree (BMT)} \cite{bmt2007} is proposed by Rogers \emph{et al.} to alleviate the high MAC storage and access overhead of MT. They use EncCtr as an additional input of MAC computing (i.e. MAC = Hash$_{key}$(Address, EncCtr, Data)). The counter would never be reused for the same address, and the tampering with any element in the MAC computing will cause the mismatch. Therefore, just make sure the EncCtrs are fresh is enough to resist replay attacks. In BMT, each leaf node is a MAC of multiple EncCtrs, which greatly decreases the overhead of integrity verification (as shown in figure~\ref{it}(b)). 

\inpnglw{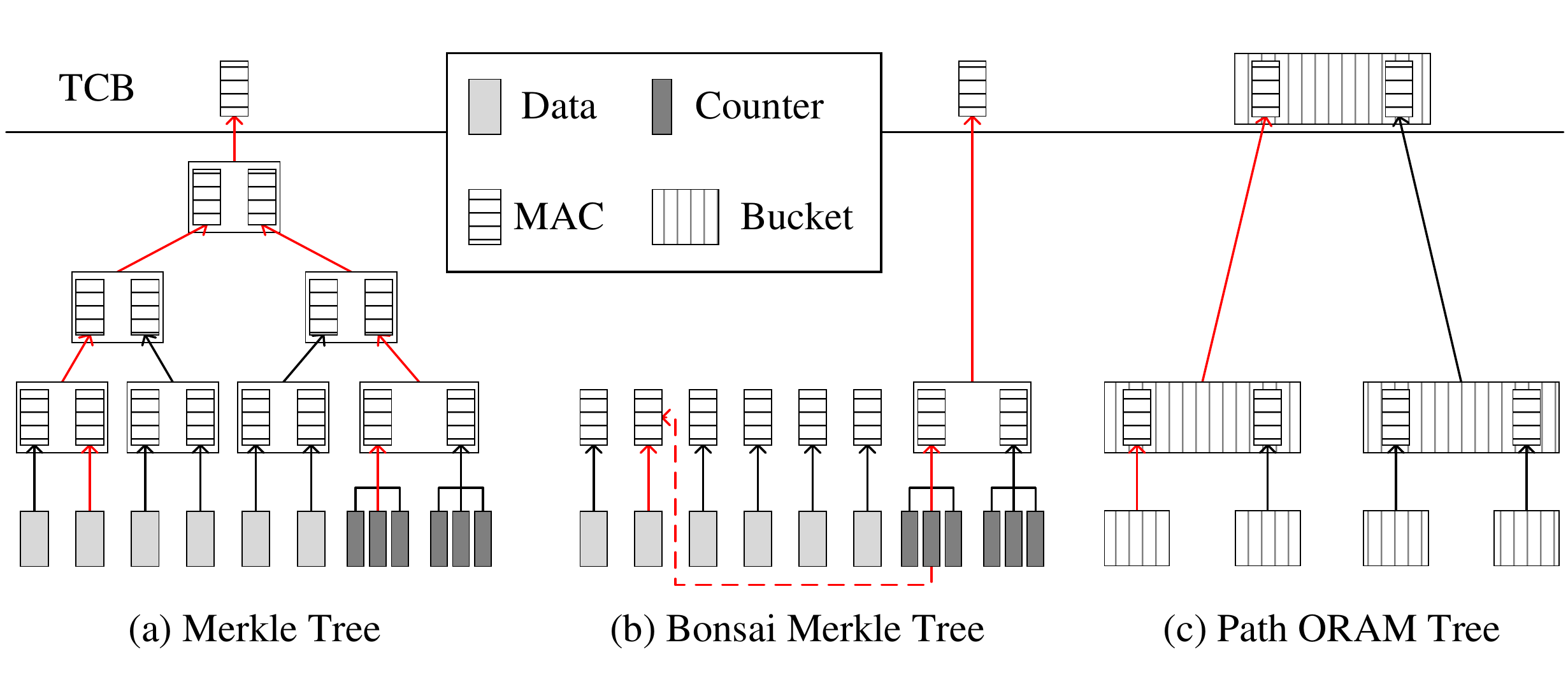}{Integrity trees for replay attacks: (a) Merkle Tree, (b) Bonsai Merkle Tree, (c) Path ORAM Tree with integrity verification.}{it}{0.48}

\textbf{Path ORAM integrity tree (PIT)} \cite{pathit2013} exploit the common operation in MT and Path ORAM to reduce the expensive integrity tree traverse. Each memory request in both MT and Path ORAM needs to read a path in their tree structure. This common path access operation can be integrated into one path access. As figure~\ref{it}(c) shows, each bucket in the ORAM tree contains not only its own data but also the MAC values of its child buckets. Similar to MT, the MAC value of the bucket is also depends on its two child buckets, and the root is stored on the TCB. Thus, the integrity of Path ORAM can be guaranteed with PIT. In a normal Path ORAM access, all blocks in a path are fetched to get the target, and then the blocks in the stash are evicted back to the path. Therefore, accesses to the metadata for integrity verification can be integrated into normal Path ORAM access and without additional memory transactions.

\subsection{Reliability Guarantee}
Memory failures are a common form of reason that results in computing systems crash. The real-world testing by Sridharan \emph{et.al.} \cite{studydramfailures2012} evaluates the failure probability of DRAM devices (see in Table~\ref{failure}). The result shows that 0.066 failures happen per billion device hours per Mbit (FIT/Mbit), the large granularity failure is as common as the single bit failure, and the permanent fault is the most prevalent cause of failures.

\begin{table}[!h]
	\centering
	\caption{DRAM FAILURES PER BILLION HOURS (FIT)\cite{studydramfailures2012}}
	\label{failure}
	\begin{tabular}{|l|c|c|}
		\hline
		\multirow{2}{*}{Failure Mode} & \multicolumn{2}{c|}{Failure Rate (FIT)} \\ \cline{2-3} 
		& Transient         & Permanent         \\ \hline
		Single-bit                              & 14.2              & 18.6              \\
		Single-word                             & 1.4               & 0.3               \\
		Single-column                           & 1.4               & 5.6               \\
		Single-row                              & 0.2               & 8.2               \\
		Single-bank                             & 0.8               & 10.0                \\
		Multi-bank                              & 0.3               & 1.4               \\
		Multi-rank                              & 0.9               & 2.8               \\ \hline
	\end{tabular}
\end{table}

\textbf{SECDED Error Correction Code (ECC) Memory}\cite{eccmemory1970} is wildly used in the compute clusters where undetected data corruption is unacceptable. It provides an additional ECC chip for every eight data chips. The ECC can be fetched parallel with the data that avoid separate memory access.
SECDED ECC can correct single-bit errors and detect double-bit errors, but the three-bit error may be mistaken for a single-bit error and the four-bit error may not be detected at all.
Therefore, more stronger error detection code such as CRC8-ATM code\cite{crc82005} and \textbf{MACs}\cite{hmac1996} are required. 
For the system that requires channel-level error resilience, redundant arrays of independent disks (RAID) based approaches\cite{mirroring2001,hp2008,ibm2012,redundant2017} are most simple and effective.
The RAID-1 based correction such as Mirrored Memory\cite{hp2008} write data and its \textbf{Replica} into two channels simultaneously. Once a channel is failed, another channel can be used to recover the failed channel.
The RAID-3 based correction such as RAIM\cite{ibm2012} using a separate memory channel to store the parity information of all the other data channels. The failed channel can be recovered through XOR the parity channel and the other correct data channels.

With the memory scaling to a smaller feature size, the cell failed rate is likely to increase to 10$^{-4}$\cite{archshield2013}. To tolerate cell damage and guarantee memory reliability and availability, different granularity failed cell repair strategies are proposed. The main idea of these works\cite{layser2003,ecp2010,archshield2013,ecs2017,flower2020} is to use the spare memory region to store the correct value of the failed memory cells. \textbf{Error Correction Pointer (ECP)}\cite{ecp2010} is a cell-level granularity fault replacement strategy. 
Each ECP stores the address and the correct value of one failed memory cell.

\section{Low Overhead Integrity Verification}\label{sec_rit}

\subsection{Re-purposing of ECC Chip}
Similar to previous security and reliability co-designs\cite{synergy2018,reducing2018,compact2020}, our secure memory system is built on the 9-chips ECC-DIMM. 
We re-purpose the 9th chip (ECC chip) to store the different types of metadata other than just ECC. 
For example, it can be used to store the MAC to provide stronger tamper-resistant/error-detection capability. It can also be used to store ECPs to increase memory reliability and availability. 
More details of using the ECC-chips are described below.
This re-purposing allows the critical metadata can be fetched parallel with the data and avoid separate memory transactions, which has a positive impact on the system performance.

\subsection{Ring ORAM Integrity Tree} 
The common design principle of different integrity tree designs is to build the hard-to-fool verification chain between the trusted node and the untrusted node. 
MAC is a strong enough linkage for the chain as the probability of a MAC-collision is negligible. 
The integrity of an untrusted node can be verified through the chain to the trusted node.
Any tampering on the untrusted node would break the chain as the computed MAC does not match the stored MAC. 
However, the traverse on the chain incurs greatly memory bandwidth overhead.
To eliminate this overhead in Ring ORAM, we propose RIT, an efficient integrity tree without any additional memory accesses. 

\textbf{RIT} builds the verification chain over the existed memory organization in Ring ORAM.
As Figure~\ref{rit} shows, we first build the verification chains for metadata blocks over the existed binary tree structure just like in the PIT. The MAC of a child metadata block is stored in its parent metadata block, and the root bucket (Bucket$_{0}$) is stored on-chip. The two MAC values in the root are computed based on all metadata blocks in the left and right ORAM subtree separately. Thus, any illegal modification in metadata blocks would cause the root MACs mismatch. In each Ring ORAM access, the metadata block and its MAC are fetched and updated together. Due to the MACs are stored in the metadata block and can be fetched in one memory block, there is no additional memory access.

\inpnglw{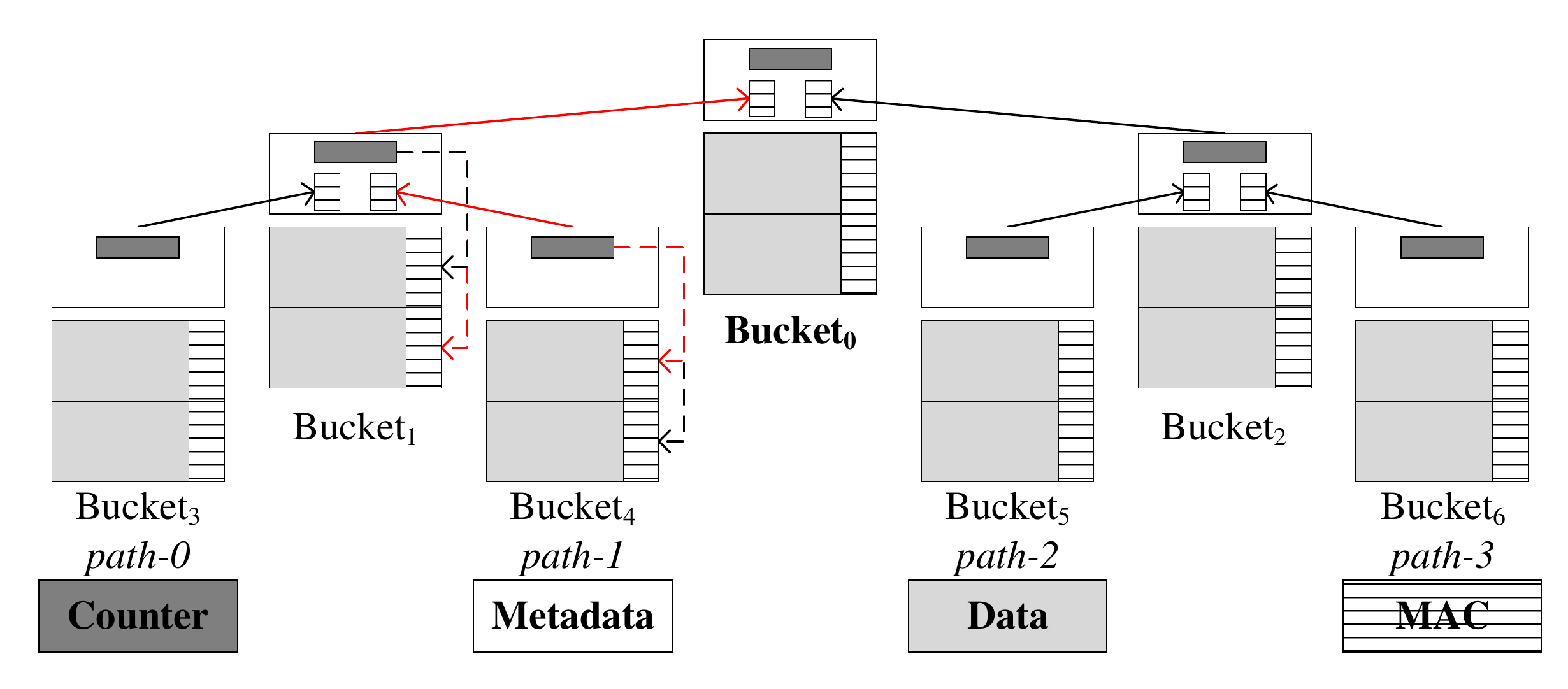}{Ring ORAM Integrity Tree (RIT).}{rit}{0.48}

Then, we build the verification chains between the metadata block and all data blocks in the same bucket. Similar to BMT, the EncCtr is used as an additional input of MAC computing for the data blocks in Ring ORAM. Ensuring the EncCtr is not tampered with is enough to verify the integrity and freshness of the data blocks with MACs. In the Ring ORAM tree, one EncCtr is shared by all the blocks in a bucket and is stored in the metadata block. Therefore, protecting the metadata block from tampering guarantees the integrity of all data blocks in the bucket can be verified. To avoid additional memory access for the MAC, we store it in the ECC-chip of the data block.

By building the verification chains between metadata blocks and between the metadata block and corresponding data blocks, RIT ensures that any tampering within the Ring ORAM tree can be detected.

\textbf{Integrity Verification Procedures.} Three main operations in Ring ORAM with RIT adopted is described as follows. The examples are based on Figure~\ref{rit}. Let B$_{i}$ be the \emph{i}-th bucket in the ORAM tree, MB$_{i}$ be the metadata block of \emph{i}-th bucket. The root bucket B$_{0}$ is cached on-chip.

\emph{Read Path}: Assuming the target block is located on \emph{path-1}. 
\circled{1} Metadata blocks are read from buckets (B$_{1}$, B$_{4}$) on \emph{path-1}. Then, we check whether the Hash$_{key}$(Address, MB$_{i}$) is equal to the MAC stored in its parent node (MB$_{(i-1)/2}$). If all is equal, the metadata blocks on \emph{path-1} are correct and fresh. Of course, the EncCtr is correct.
\circled{2} One data block with its MAC are fetched together from each bucket (B$_{1}$, B$_{4}$) on \emph{path-1}. If the Hash$_{Key}$(Address, EncCtr, Data) is equal to its MAC, the data block is ascertain not tampered with. 
\circled{3} Update VBits and ReadCtr in the metadata blocks (MB$_{1}$, MB$_{4}$). Re-computing the MAC of MB$_{i}$ (MB$_{4}$, MB$_{1}$) and overwrite it to the corresponding MAC storage space in its parent metadata block (MB$_{1}$, MB$_{0}$).

\emph{Evict Path}: The metadata blocks and the \emph{Z} blocks per bucket on the eviction path are read and verify the integrity just as in the \emph{Read Path}. Then, the encrypted stash blocks with corresponding MACs are evicted and written to the path following the eviction rules. Last, the metadata blocks are updated from the leaf to the root. 

\emph{Early Reshuffle}: Perform the read metadata block and data blocks, integrity verification, and evict operations in \emph{Eviction Path} only for the target bucket. Then, the MAC value of the metadata blocks from the target bucket to the root is updated. 

\subsection{Minimum Update Subtree Tree} \label{sec_must}
On every Ring ORAM access, the metadata block of the accessed bucket must be updated. This contributes a third of memory transactions per Ring ORAM access. 
But, we observe that only the VBits and ReadCtr are updated that only occupy 2.6\% storage space of the metadata block. 
In order to reduce the number of memory blocks that need to be operated, we separate VBits+ReadCtr out of the metadata block and organizes the consequence of them as a subtree organization. 
For example, in Figure~\ref{must}, we separate the VBits+ReadCtr sets from a 6-level ORAM tree and construct them as a binary tree (called PMetaTree). 
Then, the PMetaTree is separated into multiple 3-level subtrees, and each subtree is one memory block size (i.e. 72-byte in ECC memory). 
We call such subtree as Minimum Update Subtree (MUS). 
This can reduce the number of blocks required updated from 6 to 2.

\inpnglw{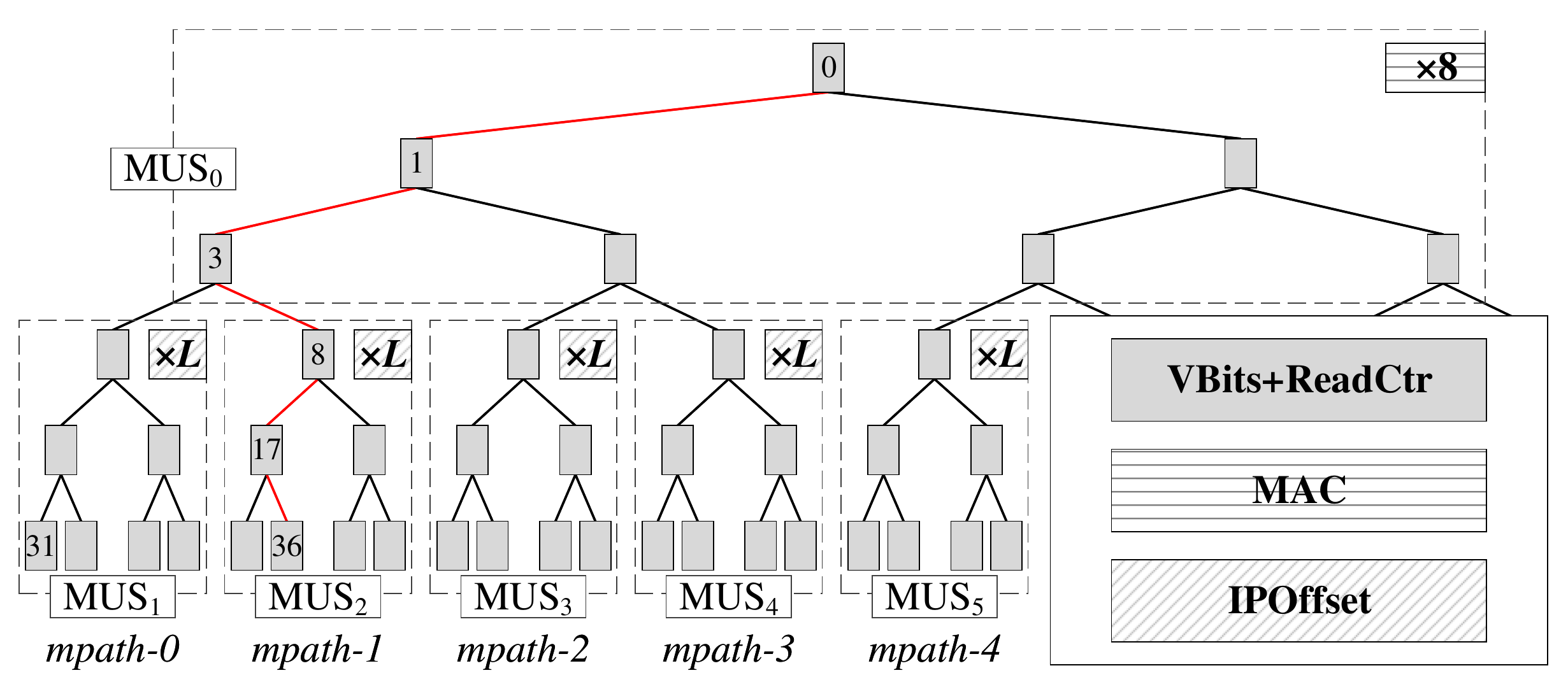}{Minimum Update Subtree Tree (MUST).}{must}{0.48}

\textbf{Minimum Update Subtree Tree (MUST)} is constructed by the MUS, and each MUS is a node in the MUST. 
When we want to access a path in the PMetaTree, e.g. the red path in Figure~\ref{must}, we first read multiple MUST nodes (i.e. MUS$_{0}$ and MUS$_{2}$) that contains the target path. 
Then, we check the integrity of these nodes. 
Next, to get the target metadata, we need to calculate the internal target path in each accessed MUST node (e.g. 0-1-3 in MUS$_{0}$). 
After the VBits+ReadCtr sets are used and updated, we re-computing the MACs and write all these data back to the MUST.

The target path in the MUST can be easily calculated through the path label in PMetaTree and the leaf node number in the leaf MUST node. Using the red path in Figure~\ref{must} as an example. The start leaf node label is 31 and the target path label is 36, which means the target path is from the 6th leaf node of PMetaTree to the root. And each four leaf nodes in PMetaTree belongs to a leaf MUST node. Therefore, the target path in MUST is from the $\lceil6/4 \rceil$-nd leaf MUST node to the root, i.e. \emph{mpath-1}. The internal path in the $\lceil6/4 \rceil$-nd leaf MUST node is easy to determine, as the node-36 is the second leaf node (also the 5th node) in the MUS$_{2}$, thus the internal path is 8-17-36 (i.e. the 1st, 2nd, 5th node of MUS$_{2}$). 

To reduce the internal path calculation overhead of the non-leaf MUST node, we use the Internal Path Offset (IPOffset) to mark the internal target path. Using the MUS$_{0}$ in Figure~\ref{must} as an example, the IPOffset of MUS$_{0}$ on \emph{mpath-1} is 3 that means its internal path is from the 4th node to the root when we access \emph{mpath-1}. As the IPOffset of non-leaf MUST nodes on a path is determined by the leaf MUST node, we can store all these IPOffsets into the corresponding leaf MUST node.

The integrity protection of MUST is similar to PIT, each non-leaf MUST node holds MACs of its eight children. One path access can fetch all the required data and MACs.

\subsection{Security Analysis}
Our proposal does not change the features and operations in Ring ORAM and therefore does not influence confidentiality and obliviousness.
The RIT follows the design principle of integrity tree and can protect memory integrity and freshness from replay attacks. 
The MUST separate a part of plaintext format metadata from the ORAM tree to reduce the update overhead, which does not influence security.
To protect MUST from replay attacks, we applied a PIT-like scheme for it.

\section{Channel-Level Error Resilience}\label{correction}
\subsection{Secure Replication for Ring ORAM tree}

As the ECC chip is occupied by the MAC, we need additional memory access to fetch metadata to correct errors.
In previous testing\cite{studydramfailures2012}, the memory failure rarely occurs compared with the number of memory operations, i.e. one failure per six hours across the Jaguar system. Most of the time, the ECC is served for error detection other than correction. Therefore, the program slowdown caused by correcting errors after MAC mismatch can be ignored. A similar conclusion can also be obtained from \cite{ivec2010}. What is more important is to correct large granularity errors which have been proven as frequent as the single-bit error.

Data replication is one of the most straightforward and reliable methods to solve this problem. 
Errors can be corrected as long as the data and its replica do not have faults at the same time.
However, it incurs heavy storage overhead, at least half of the storage is wasted (with one backup). Fortunately, Ring ORAM can completely reduce this overhead with the dummy slot.

The dummy slot, which stores meaningless data, could be used to store the replica of data. The replicas are served as dummy blocks and dropped directly in normal Ring ORAM operations that will not cause data inconsistency or security compromise. When the failures occurred and be detected, the replicas in the bucket can be fetched to correct errors. In the rest of this paper, we use five real slots and seven dummy slots per bucket as the default parameters as they are the most efficient setting for Ring ORAM in secure processors\cite{ring2015}. 

\textbf{Reliability Aware Replication.}
Simply replicate blocks from the real slots to the dummy slots cannot tolerate device-level failures, such as chip failures and channel failures. Previous study\cite{eccparity2014} shows that as channels are independent of each other, failures typically occur in only one channel at a time. This motivates us to store the original real block and its replica in different channels to gain a channel-level error resilience. In this paper, we assume a two-channel memory system as the default. Following strategies are used to provide channel-level error resilience for the Ring ORAM tree:

\emph{Strategy 1: Evenly address mapping policy.} Each bucket in the ORAM tree is mapped to physical memory space follows the order of "row:bank:column:rank:channel:offset". So that blocks in the bucket are evenly mapped to different channels.

\emph{Strategy 2: Reliability aware replication.} After determining the location of real blocks in a bucket, the replicas \emph{could} and \emph{must} be stored in different channels from their corresponding original blocks. So that even an entire channel is inaccessible, the lost data still could be recovered.

\emph{Strategy 3:  In-order replicas storage.} The location of real blocks in the bucket is identified with the \emph{offset}.  
To eliminate the offset storage overhead of replicas, an in-order storage strategy is used to fill the replicas into the dummy slots. Firstly, the replicas are arranged with the order of the \emph{addresses} stored in the metadata. Then they are filled into the dummy slots from left to right in sequence of the order under the limitation of \emph{Strategy 2}. With this strategy, the location of replicas can be calculated by the addresses and offsets in the metadata. 

Figure~\ref{replication} shows the steps of the Reliability Aware Replication scheme. First of all, the slots in the bucket are evenly mapped to two channels and each channel holds six data slots. After the location of real blocks is determined, the replicas are filled into the dummy slots. The replica of the metadata block is the first block to fill (see Figure~\ref{replication} (a)). Due to the metadata block (i.e. M) is stored in \emph{channel-1}, the first dummy slot in \emph{channel-0} is searched from the left to the right to fill the replica (i.e. m) of the metadata block. Then, follows the order (i.e. A, B, C, D, E) in the metadata to create replicas and store them in the dummy slots. As Figure~\ref{replication} (b) shows, the replica (i.e. a) of block-A is filled into the first empty dummy slots in \emph{channel-1}. Similarly, the replicas of B, C, D, E are filled in sequence as Figure~\ref{replication}~(c)~(d) shows.

\inpnglw{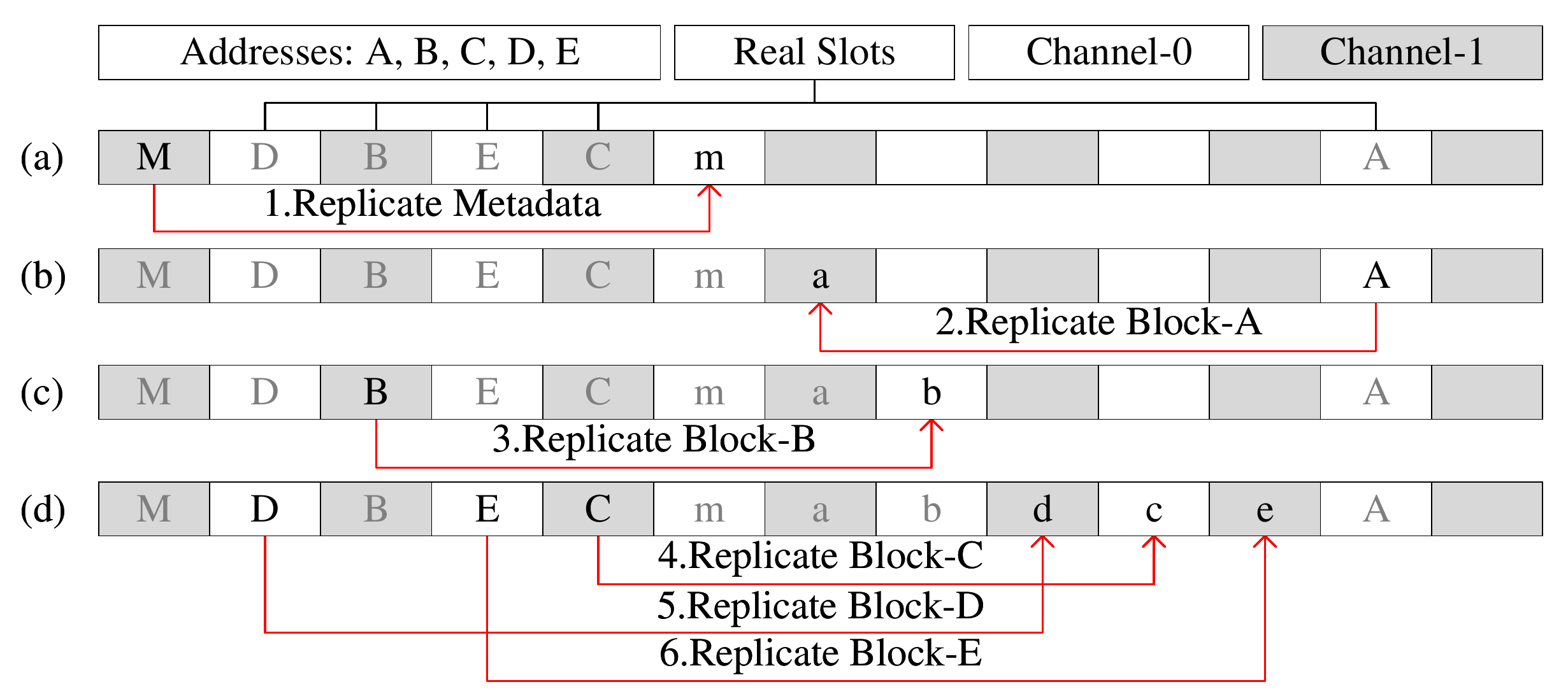}{An illustration of Reliability Aware Replication.}{replication}{0.48}

\textbf{Secure Correction.}
We assume that failure occurs in only one channel at a time. After integrity verification failed, the location of the error block is identified. Then, we would fetch the replicas to correct errors. To avoid the real slots and dummy slots being distinguished, all blocks in the same bucket as the error block but in different channels are fetched to the processor and checked for integrity. 
Next, we will use three cases to describe the details of the error correction method. 

\emph{Case 1: Real blocks / Replicas error.} We use the offset of real blocks and the \emph{In-order replicas storage} strategy to get the location of real blocks and replicas. Then, the error block can be corrected with its replica.

\emph{Case 2: Metadata Block error.} When we want to use the replica to correct the error in the metadata block, there are two challenges to be solved. The first one is the encrypted blocks maybe failed to decrypt as the EncCtr maybe failed. The second one is the location of the replication metadata block maybe not known as the \emph{offsets} may be lost. 

\emph{Partial EncCtr.} To address the first challenge, we need to store the replica of EncCtr (60-bit size) in other slots in plaintext format. But this 60-bit replica is too large to be stored in one data slot. Therefore, we divide the EncCtr into six 10-bit Partial EncCtr and spread them across multiple data slots on the same channel. The 64-bit ECC area of data slots stores 54-bit MAC and 10-bit Partial EncCtr. For each bucket, 12 data slots are evenly mapped to two channels and can hold two replicas of EncCtr. Thus, even if a channel is failed, the correct EncCtr can still be recovered.

After we reconstruct the EncCtr through the Partial EncCtrs and correctly decrypt all the blocks. We can address the second challenge by trying each fetched block as a possible replica of the metadata. The MAC stored in its parent metadata block can be used to check the correctness of the tried data. Once the integrity verification is passed, the metadata is corrected and the subsequent MAC computing will be terminated.

\emph{Case 3: Channel Failed.} The difficulty of channel failed recovery still lies in the recovery of the metadata block.
In such a failure case, the parent and child metadata blocks are likely to be failed simultaneously. This causes the MAC in the parent cannot be used to check the correctness of the possible replicas of the child metadata block. 
Fortunately, as the MACs in the root buckets are stored on-chip are fault free, we can check the integrity from the root to the leaf recursively. 
Like in the \emph{Case 2}, we first use the reconstructed EncCtr and the MAC stored in the fault-free/corrected metadata block to recovery the child metadata block.
Then, the location of real blocks/replicas is calculated through the offset and the storage strategy. 
Last, the recovered data are filled into the new replacement DIMM.

\subsection{Mirrored Minimum Update Subtree Tree}
The main idea in Secure Replication is to make full use of the wasted storage space in the Ring ORAM tree. This idea cannot be extended to a compact data organized MUST. We have to allocate an independent area to store the replica of MUST. We call such a replication scheme Mirrored MUST (MMUST).
Its structure is shown in Figure~\ref{mmust}.
Each node in the MUST has a replica in a different channel, and one MUST is built on two channels to improves memory channel utilization.
In each ORAM access, a path in the MUST is read to get useful information, then the updated metadata is written to two MUST simultaneously to ensure data consistency. 

\inpnglw{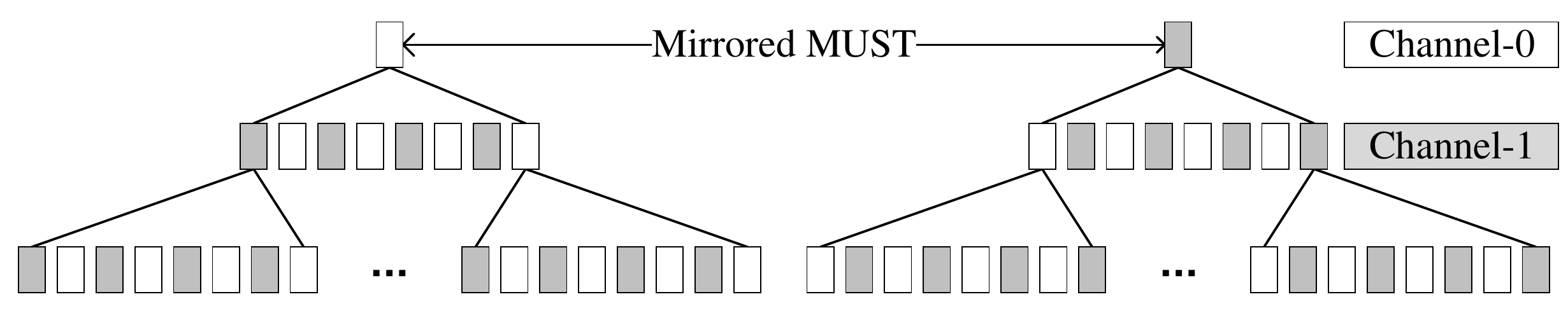}{Mirrored Minimum Update Subtree Tree (MMUST).}{mmust}{0.48}

Traditional mirrored memory techniques\cite{mirroring2001,hp2008} are cost-prohibitive to deploy, except applying for a smaller memory region (e.g. MUST). The size of a MUST is much smaller compared to the Ring ORAM tree. Assuming an ORAM tree with 6GB capacity, 23 tree levels, and the top seven levels cached. The 15-bit VBits+ReadCtr set is separated from each bucket in the ORAM tree to construct a 5-level MUST that requires 20.57 MB space. The capacity of the Mirrored MUST is only 0.667\% of the Ring ORAM tree and is acceptable. 

Since the top 2 levels of the MUST only occupy 40.5 KB storage space, we can store them in the cache to decrease the height of MUST. Under such a setting, the number of memory block operations in each ORAM access is decreased from 48 (32 reads and 16 writes) to 41 (35 reads and 6 writes). Even with the access to replicas, the memory access overhead of our scheme is still lower than the Ring ORAM without MUST.

\subsection{Security \& Reliability Analysis}
\textbf{Security Analysis.} Secure Replication scheme does not change the memory access pattern in Ring ORAM in normal operations. When a failure occurs, with the over-fetch of blocks, the real slots and dummy slots are still indistinguishable. Therefore, the Secure Replication scheme does not reduce ORAM security. The content in the MUST is public information that no need to guarantee its confidentiality. The integrity of the mirrored MUST nodes is also protected with the integrity tree.

\textbf{Reliability Analysis.} With our Secure Replication and Mirrored MUST schemes, any failure that occurs in one channel can be corrected. The probability of errors occur in the same relative locations in different channels is too small to happen. Following the assumption in ECC Parity\cite{eccparity2014}, there is only 0.002 chance that two or more channels develop faults in any single detection window during seven years of operations.

\section{Permanent Faults Repair}\label{repair} 
The permanent fault is the most prevalent cause of failures\cite{studydramfailures2012}.
Although the errors can be corrected by the replicas, leaving the fault memory cells still service reduces both memory reliability and system performance.
In this paper, we assume the fault incidence rate of the DRAM memory cell is 10$^{-4}$, which is the tolerable limit for many existing schemes\cite{archshield2013,xed2016,flower2020}.
Error Correction Pointer (ECP) can be used to repair the permanent fault by replacing the fault memory cells with spares. But similar to other fault replacement strategies\cite{archshield2013,flower2020}, it also incurs additional memory overhead. When a data block is fetched from memory, additional fault repairing information is accessed to judge and correct the fault. 
To reduce the ECP storage and access overhead, we integrate it into the structures in Ring ORAM and MUST.

\subsection{Permanent Faults Repair for Ring ORAM Tree}
In Ring ORAM, the metadata block of each bucket is aligned to one memory block size, which causes some storage space wasted.
We can store the permanent fault repair information in this space to avoid additional memory overhead.
The ECP in the metadata block can be used to repair the faults in the whole bucket.
Although ECPs can repair more failed cells than ECC with the same storage overhead, they cannot repair the faults that happen in themselves. A careful design is needed to address this problem.

When a permanent fault happens in an ECP, another ECP would be used to repair it. An in-sequence repair scheme is important to avoid the loop pointer, i.e. two ECPs are point to each other that causes both to fail. Thus, in the repair scheme, we guarantee the rule that \emph{a fault ECP must be repaired by the ECPs in front of it}. 
Simple examples are given in Figure~\ref{iecp}. We assume that an ECP would have at most one cell failed. 
As shown in Figure~\ref{iecp} (b), the permanent fault in ECP-3 is repaired by ECP-1.
When the region where the first ECP (ECP-1) located suffered permanent faults, we rotate ECPs left one ECP size to keep the first ECP fault free (see Figure~\ref{iecp} (c)). We use rotate offset (ROffset) to record the distance that ECPs are rotated. After the fault in ECP-2 is repaired by ECP-1, the ECP-2 can be used to repair another fault (see Figure~\ref{iecp} (c)). To avoid the loop pointer, at most four permanent faults can be tolerated in the region with five ECPs (see Figure~\ref{iecp} (d)). Similarly, when an ECP has two cells failed, we must reserve two correct ECP to repair it. If there are not enough correct ECPs before it, we also need to rotate the ECPs.

\inpnglw{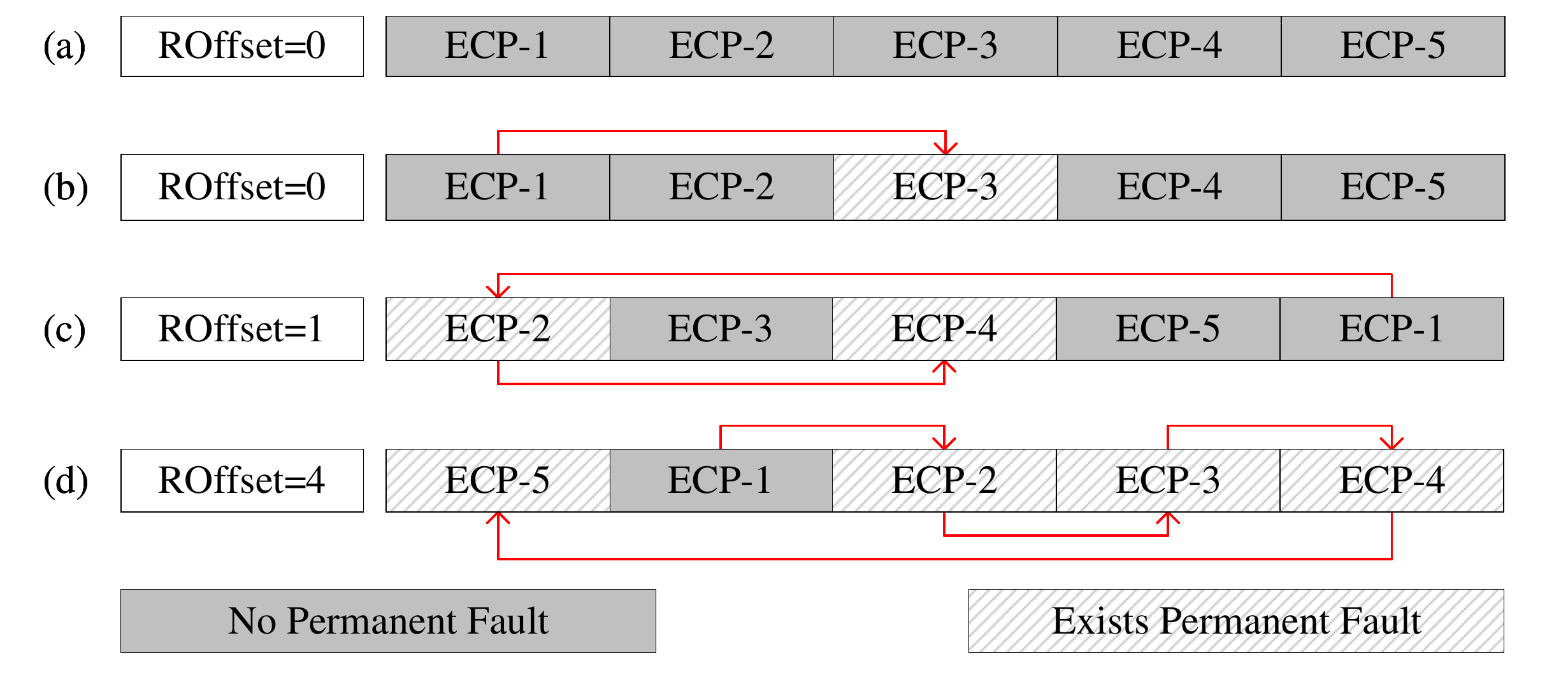}{An illustration of repairing the permanent fault in the ORAM bucket.}{iecp}{0.48}

Another situation that must be considered in our work is the permanent fault in the replicated ECPs (called RECPs). For example, if ECP-1 is used to repair the fault in its replica RECP-1, once the metadata block is failed, the RECP-1 cannot correctly repair itself. Therefore, before filling the bucket, we first find a slot without permanent fault in the same location as the ECPs in the metadata block to fill the replicated metadata. The choice of the slot is also restricted by the Reliability Aware Replication scheme. We use an additional offset label in metadata to mark the position of the replicated metadata.

We rely on the MAC to detect errors. After an error is detected and corrected, the correct data is written back and read again to check the type of fault. If the error is caused by the permanent fault, we execute \emph{Early Reshuffle} to update the ECPs in both the metadata blocks and the replicas.

\subsection{Permanent Faults Repair for Mirrored MUST}
In the Mirrored MUST, the ECPs can be co-located with the MUST node to avoid separate memory accesses. Because the node and its mirror have the same data, the ECPs are required to repair the permanent fault in both the node and its mirror. Therefore, we also need to guarantee that an ECP would not point to itself and its replicas (RECP).

Similar to the repair scheme in the Ring ORAM tree, we guarantee that a fault ECP/RECP must be repaired by the ECPs/RECPs in front of it. Examples are given in Figure~\ref{mecp} to present the detail of the repair scheme in Mirrored MUST. As shown in Figure~\ref{mecp} (b), each failed cell can be repaired by both an ECP and a RECP. When a permanent fault happens in the first ECP region, we need to rotate ECPs and RECPs left at the same distance so that the ROffset in both nodes can be kept the same. When a hard fault happens in the RECP-1, we also need to rotate both ECPs and RECPs to keep the RECP-1 fault free and ROffset the same (see Figure~\ref{mecp} (c)). Additionally, when both the ECP and its RECP have one cell failed as Figure~\ref{mecp} (d) shown, we need to reserve two correct ECP/RECP before ECP-3/RECP-3 and repair it.

\inpnglw{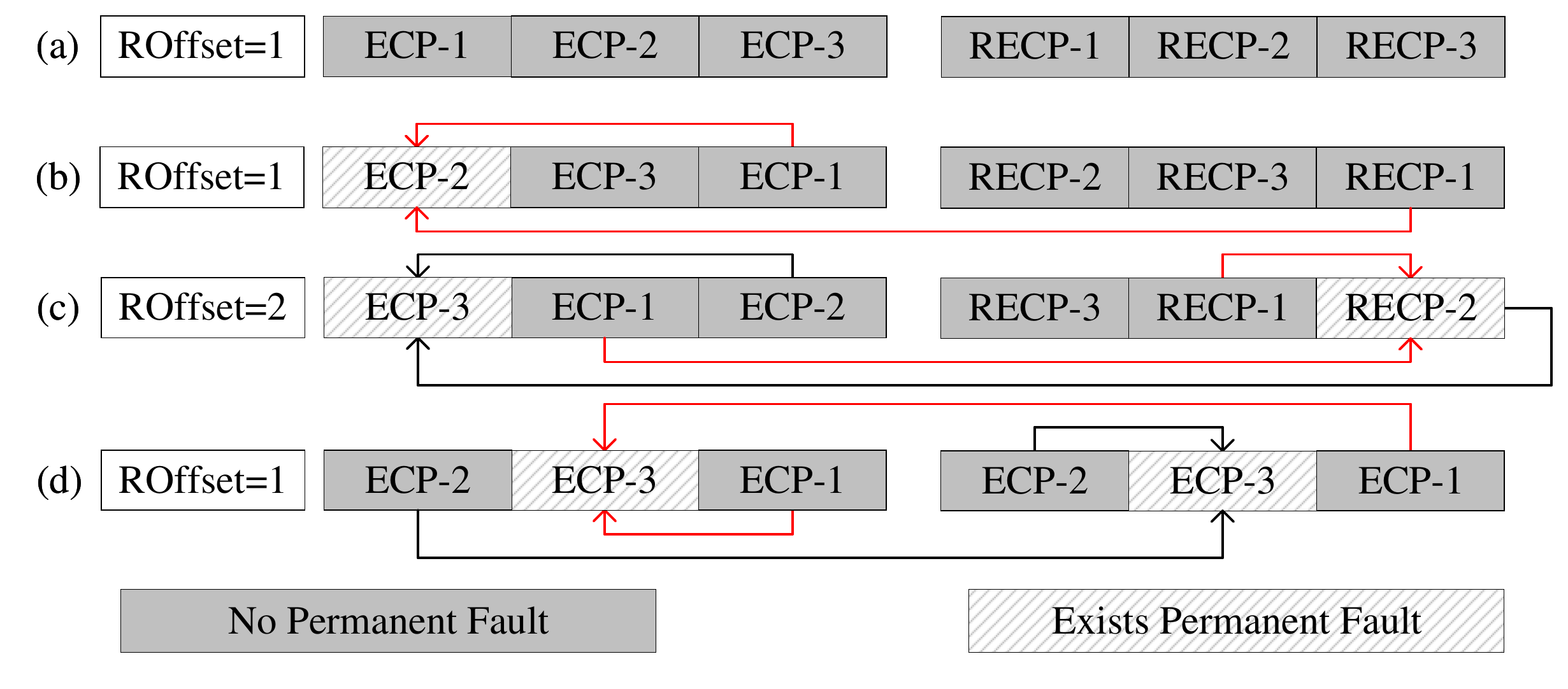}{An illustration of repairing the permanent fault in the MMUST.}{mecp}{0.48}

To detect and correct the permanent faults in Mirrored MUST, we let the MUST and the Mirrored MUST service for data reading in turn. We rely on the MAC to detect errors and repeatedly write and read to detect permanent faults.

\subsection{Security Analysis}
The permanent fault information is not sensitive and does not need confidentiality protection. 
The ECP is also protected by the integrity tree and cannot be tampered with.
Therefore, the repair method would not cause security problems.

\section{Compact Metadata Design}\label{compact}

For the Ring ORAM in secure processors, metadata access is a non-negligible overhead. To avoid the over-fetch of metadata blocks, we must carefully design the metadata structure to let them can be fetched in one memory block (i.e. 72-byte in ECC-DIMM based memory systems).

\subsection{Metadata Design for Ring ORAM Tree.}
Figure~\ref{icm} shows the compact metadata design in the Ring ORAM tree. The gray part is the data that needs to be protected by encryption.
Figure~\ref{icm} (a) shows the data organization of the bucket. Each bucket has one metadata block, at most five real blocks and at least seven dummy blocks, the location of the data blocks are randomly permuted (not shown in the figure for simplicity). Six of the dummy blocks are replaced by replicas of the metadata block and real blocks. The ECC chip of data blocks stores a 10-bit partial EncCtr and a 54-bit MAC.  

\inpnglw{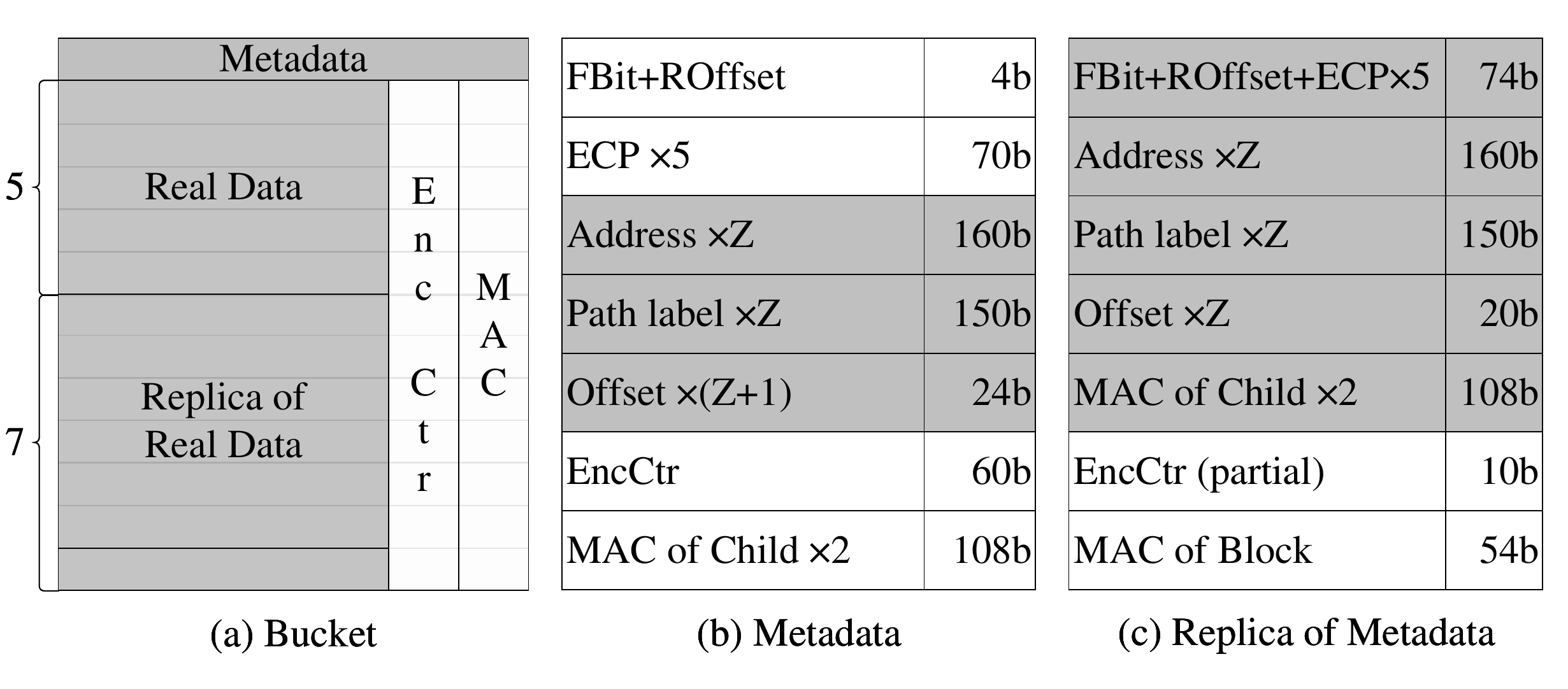}{Compact metadata design in the Ring ORAM tree.}{icm}{0.48}

Figure~\ref{icm} (b) shows the compact metadata organization. One fault bit (FBit) to note whether the metadata block suffers permanent faults. 3-bit rotate offset (ROffset) is used to record the rotate distance of the five ECPs. Each ECP uses 13-bit to store the fault cell address in the 7588-bit bucket, and 1-bit to store the correct value of the fault. 32-bit address and 30-bit path label space are reserved for \emph{Z} real blocks. An additional 4-bit offset is used to record the address of the replicated metadata block. 60-bit EncCtr for encryption/decryption. Two 54-bit MACs belong to its two child metadata blocks.

Figure~\ref{icm} (c) shows the organization of the replicated metadata block. There is a little difference between the metadata block and its replica. The replica does not need to have the offset of itself in the bucket, and the EncCtr is useless when it is encrypted. The replicas and the other blocks expose the same data structure to attackers to make them indistinguishable, i.e. 512-bit encrypted data, 10-bit partial EncCtr, and 54-bit MAC. The MAC is used to verify the integrity of the 512-bit encrypted data and 10-bit partial EncCtr.

The ECPs in a bucket can repair at most four failed cells in the ECP region, and five failed cells in the total bucket. We term the failed cell count exceeds the repair capability of the ECPs as ECP failure. The probability of the failure is 1.6$\cdot$10$^{-13}$ for the former and 1.29$\cdot$10$^{-4}$ for the latter. This means more than four cells are failed in one ECP region may never happen. When a bucket has more than five cells failed, the whole bucket would be remapped to a redundant area. For a 6GB Ring ORAM tree, there may be at most 1084 buckets that need to be remapped which require an 8KB remapping table. We can cache it on-chip to reduce its consulting overhead.

\subsection{Metadata Design for MUST.}
Figure~\ref{mcm} shows the metadata design of the MUST node. The leaf node and the non-leaf node have different compositions.
The non-leaf MUST node has one FBit to note whether the node and its mirror has a permanent fault, one 2-bit ROffset for three 12-bit ECPs, seven 15-bit VBits+ReadCtr sets, eight 54-bit MACs that belong to its eight child nodes (see Figure~\ref{mcm} (a)). Without the MAC storage overhead, the leaf MUST node contains 31 sets of VBits+ReadCtr, seven 12-bit ECPs, and one 3-bit ROffset (see Figure~\ref{mcm} (b)). Each IPOffset is 3-bit size, and the number of IPOffset in each leaf MUST node is depended on the hight (\emph{L}) of the MUST.

\inpnglw{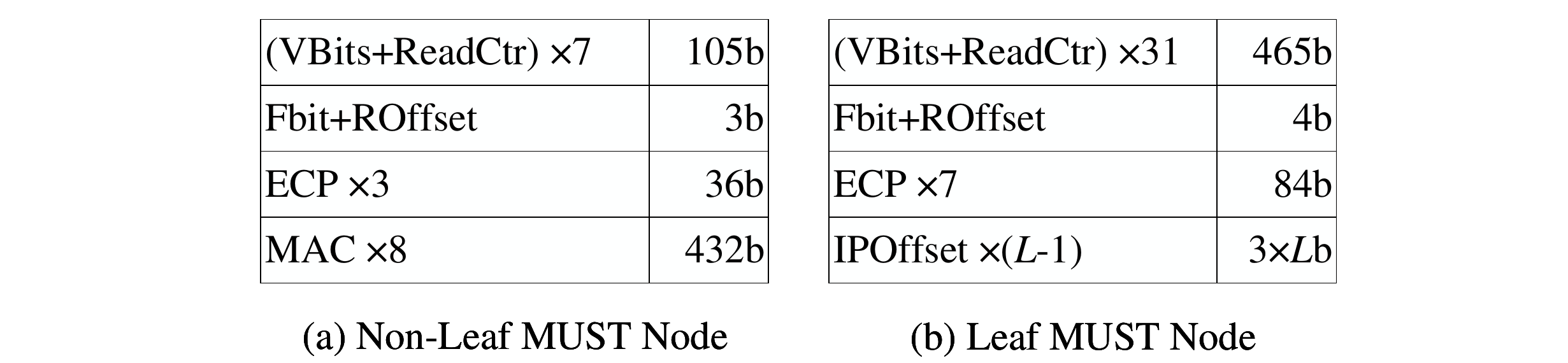}{Compact metadata design in the MUST.}{mcm}{0.48}

Table~\ref{ecpfailure} shows the probability that the ECPs cannot repair all failed cells in the MUST node. 
As the leaf MUST node provides seven ECPs, the probability that ECPs failed to repair all the faults is negligible. 
For the non-leaf MUST node, the probability of three ECPs cannot repair the faults in ECPs/RECPs region is a hundred lower than that more than three cells failed in the two nodes (a MUST node and its mirror). 
Thus, we only consider the situation that more than three cells failed in the non-leaf node and its mirror. 
For a 6GB Ring ORAM tree, the number of non-leaf MUST node is 36864. 
The expected number of the failure is 0.25 which means such failure may also never happen.

\begin{table}[!h]
	\centering
	\caption{PROBABILITY of ECP FAILURE}
	\label{ecpfailure}
	\begin{tabular}{|l|c|c|}
		\hline
		The Reason of ECP Failure      & Non-Leaf Node                & Leaf Node                     \\ \hline
		Failed in ECPs/RECPs region & 5.9$\cdot$10$^{-8}$ 		 & $\approx$0                             \\
		Failed in node/mirror node  & 6.7$\cdot$10$^{-6}$          & 6.8$\cdot$10$^{-13}$ \\ \hline
	\end{tabular}
\end{table}

\section{Experiments}\label{sec_exp}

\subsection{Methodology}
To evaluate the performance of our scheme, we conduct cycle-accurate simulations with workloads selected from SPEC\cite{spec2006} and PARSEC\cite{parsec2008}. 
We generate memory trace for each selected workload by running 500 million instructions on GEM5\cite{gem52011} full system simulator.
These traces are then fed into USIMM\cite{usimm2012} for cycle-accurate memory system simulation. 
Two memory channels are employed and the default memory timing parameters from USIMM are adopted in the evaluation. The detail of configurations is shown in Table~\ref{sc}.

We modify the USIMM simulator to implement our design and other ORAM schemes. 
The ORAM tree is 23 levels and the top seven levels are cached on-chip. 
The size of the stash is set as 512KB. When the stash is nearly full, we perform additional dummy accesses and evictions. 
For the Ring ORAM tree, each bucket has five real slots for real blocks and seven dummy slots for replicas and dummy blocks. 
The MUST is five levels eight-arity tree with the top two levels stored on-chip. 
For Path ORAM and Tiny ORAM, we assume five real slots per bucket. 
To have an acceptable storage overhead in all the ORAM schemes, we assume an 80\% real slots utilization which is proven to still have reasonable performance\cite{designspace2013}. 
To improve the eviction quality, the path eviction order in both Ring ORAM and Tiny ORAM is determined by the \emph{reverse-lexicographic order} which is suggested in \cite{ring2015,tiny2015}. 

We use the AES-CTR encryption in the basic ORAM implementation. 
We assume the OTP generation can be overlapped with the memory access, and the decryption can be completed immediately when the data blocks are fetched to the processor. 
AES-GCM is a derivation of AES-CTR, which additionally provides message authentication capability. We use AES-GCM for encryption/decryption and hash in our design and assume an 80-cycle MAC computing latency for one memory block. The number of AES-GCM units is set as four as a baseline for simulation, and we vary the number to analyze its influence on performance in the sensitivity analysis.

\begin{table}[!ht]
	\caption{System Configuration.}
	\centering
	
	\setlength{\abovecaptionskip}{0.cm}
	\begin{tabular}{|l|l|}	
		\hline 
		\multicolumn{2}{|c|}{\textbf{Processor}}                     \\ \hline
		Core & single core, 3.2GHZ \\ \hline
		Re-Order-Buffer & 128 entry \\ \hline 
		Fetch and retire width & 4 per cycle \\ \hline
		L1 Cache & 1/1 cycle, 64KB, 2-way, 64B block    \\ \hline
		L2 Cache & 10/10 cycles, 1MB, 8-way, 64B block  \\ \hline \hline
		\multicolumn{2}{|c|}{\textbf{Memory}}                        \\ \hline 
		Memory type & MT41J256M4 DDR3-1600\cite{ddr32006}            \\ \hline
		Memory bus frequency & 800MHZ \\ \hline
		\multirow{4}{*}{Memory organization} & 2 72-bit Channels \\ & 1 DIMM/Channel\\ & 2 Ranks/DIMM\\ & 9 Devices/Rank \\ \hline \hline
		\multicolumn{2}{|c|}{\textbf{Ring ORAM}}                          \\ \hline 
		Data block size & 64B                                \\ \hline
		ORAM tree height & Cache (7), DRAM (16)         \\ \hline
		MUST height & Cache (2), DRAM (3) \\ \hline
		AES-GCM unit & 4 \\ \hline
		AES-GCM latency & 80 processor cycles \cite{ivec2010}  \\ \hline
		Slots per bucket & 5 real slots, 7 dummy slots          \\ \hline
		Eviction rate & 5                                  \\ \hline

	\end{tabular}
	\label{sc}
\end{table}

\inpngtlw{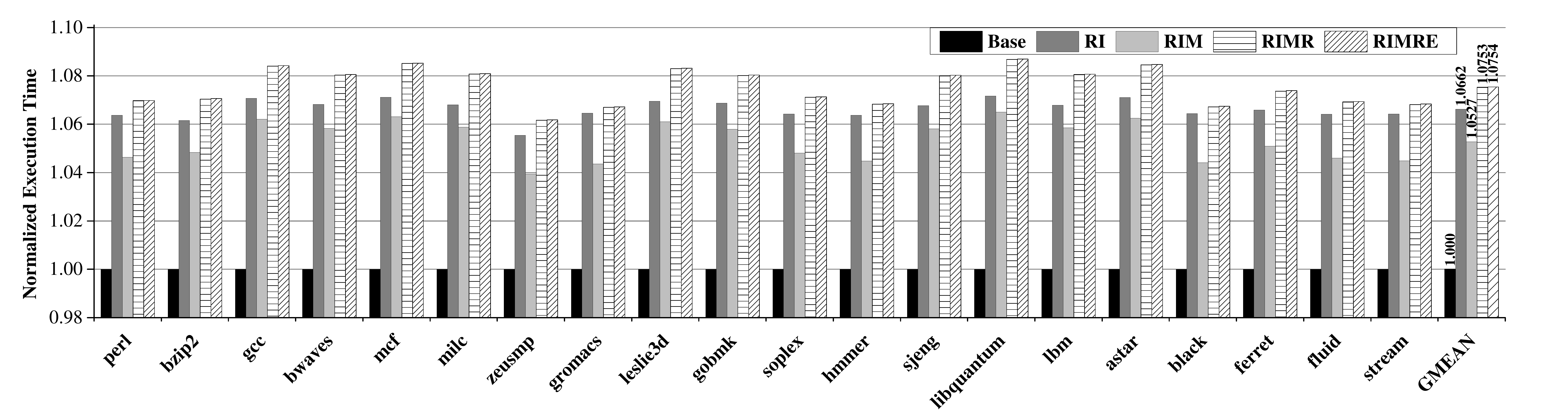}{Normalized performance of Ring ORAM, RI, RIM, RIMR, and RIMRE.}{EP}{1.04}

The following schemes are evaluated in the experiments.
\begin{itemize}
	\item Baseline. This is the original Ring ORAM implementation that used for comparison. It only provides confidential protection, no integrity verification, no reliability guarantees, and no MUST optimization. 
	\item RI: This scheme provides an integrity verification mechanism for Ring ORAM with RIT. It has four AES-GCM units that can perform parallel MAC computing. 
	\item RIM: This scheme adopts the MUST to optimize the metadata updating in RI. 
	\item RIMR: This scheme is based on the RIM. The replication and the ECPs are used to correct errors and repair hard faults respectively. We assume no error happens in RIMR.  
	\item RIMRE: This scheme is built on top of the RIMR. It considers the error correction overhead. The failure rate for modern DRAM systems is reported to be 0.066 FIT/Mbit\cite{studydramfailures2012}. For 8GB memory, this rate is translated to about an error every 8$\times$10$^8$ \textbf{seconds}, the performance overhead of correcting errors would be negligible. To make the influence of error correction more visible, we set the failure rate as an error per 8$\times$10$^6$ \textbf{processor cycles}.
\end{itemize}

\subsection{Impact on Performance}

In this section, we evaluate the performance of the proposed schemes. 
Figure~\ref{EP} shows the normalized execution time of Baseline, RI, RIM, RIMR, and RIMRE. 
From the figure, we can see that RI, RIM, RIMR, and RIMRE increase the execution time by 6.62\%, 5.27\%, 7.53\%, and 7.54\% of the Baseline respectively. 

In the RI, the MACs are fetched parallel with the data, the memory traffic of MACs is not the reason for program slowdown. Therefore, the slowdown can only be caused by MAC computing. In our setting, the processor has four AES-GCM units to compute MACs for different blocks in parallel. When there are too much data waited for integrity verification, the latter will be blocked until the previous verification complete. Increasing the number of AES-GCM units would improve performance, which will be discussed in \ref{sens_gcm}.

In the RIM, we adopt the MUST to decrease metadata update overhead, which reduces the execution time of RI by 1.35\%. 
There are three reasons that reduce the efficiency of MUST. 
The first one is the \emph{Evict Path} that performed once every five \emph{Read Path} which incurs great memory traffic. The reduced traffic of the MUST only occupy 9.19\% of the total in theoretical. 
The second one is the \emph{Early Reshuffle} that performed when a bucket is read \emph{S} times since it was last written. When the \emph{Early Reshuffle} executed, not only the node in the MUST but also the metadata blocks from the root to the reshuffle bucket need to be updated. This further reduces the efficiency of MUST. Figure~\ref{EER} shows the \emph{Early Reshuffle} percentage (\emph{Early Reshuffle} count / total \emph{Read Path} count) of 20 workloads. We can see that the average \emph{Early Reshuffle} percentage is 0.54 which means about one bucket need early reshuffle every two ORAM requests. 
The last reason is MAC computing congestion. As the MUST adopted scheme increases the number of blocks to read, there are more critical blocks waited for MAC computing. Thus, it will be helpful to provide enough AES-GCM units.

\inpnglw{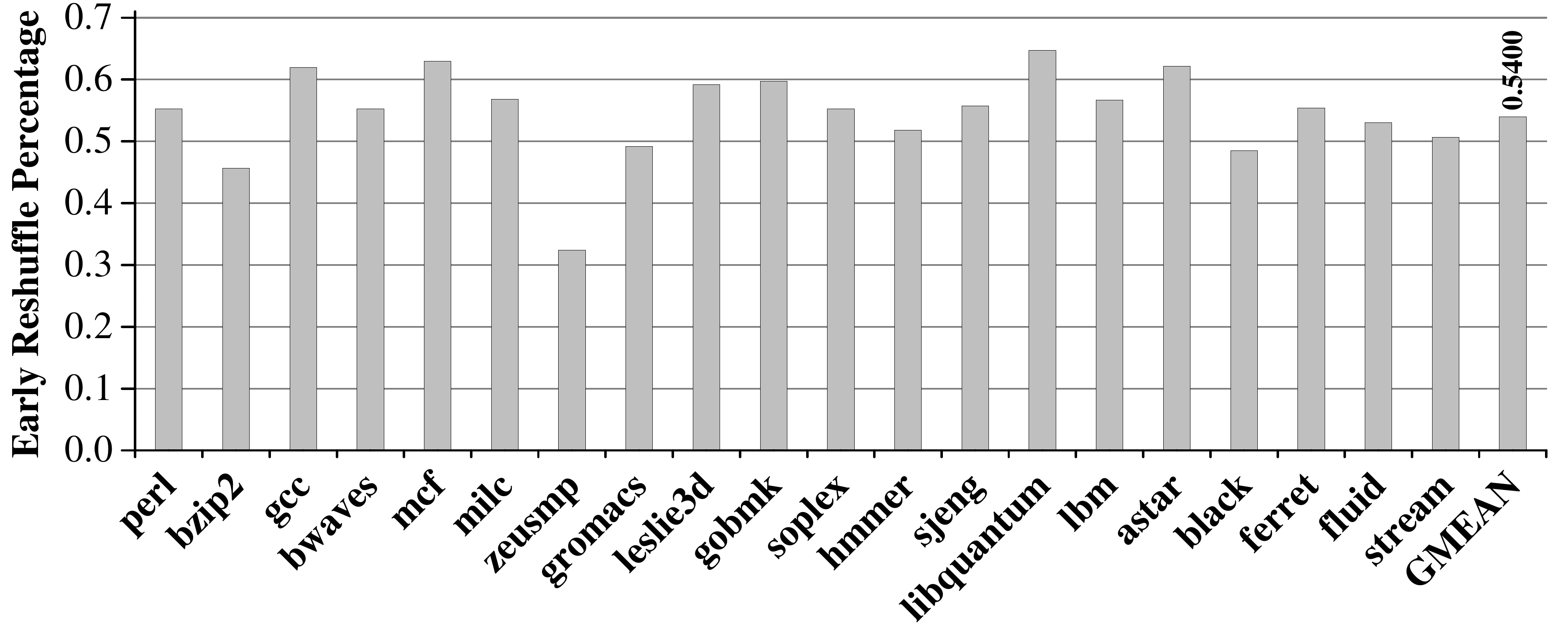}{The percentage of the \emph{Early Reshuffle} on total \emph{Read Path} accesses.}{EER}{0.48}

Although the MUST does not improve performance a lot, it is still a necessary method for our security and reliability co-design. To provide channel-level error resilience for the Ring ORAM tree, we propose Secure Replication to securely store the replica of metadata into a dummy slot. If we not using the MUST, after one bucket is accessed, the updated metadata content needs to be written back to both the metadata block and its replica. This not only incurs double metadata update traffic but also leak the location of the dummy slot that results in security loss. Therefore, separate the VBits+ReadCtr set from metadata blocks is essential in our scheme.

In the RIMR, Secure Replication and Mirrored MUST are used to tolerate completely channel failures. In the normal execution that no error occurs, the additional memory traffic comes from the write to mirrored MUST nodes. The additional reliability guarantees only increase the execution time by 2.26\% compared to the RIM.

In the RIMRE, we take the error correction overhead into the consideration. Whereas we have greatly amplified the failure rate, the overhead of error correction is still negligible, i.e. 0.01\% execution time increases compared to the RIMR.

\subsection{Impact on Memory Traffic}
In this section, we evaluate the change in memory traffic of our schemes. Figure~\ref{EMT} shows the geometric averaged read, write, and total memory traffic of the different configurations, and the results are normalized to the Baseline. RI does not require separate memory access for MACs, due to its co-location of MAC and data. RIM adopts the MUST to reduce the write traffic of RI by 14.11\%, however, additional metadata read from the MUST increase the read traffic of RI by 5.96\%.  In the RIMR, the extra reliability guarantee methods result in 5.46\% write traffic increases than RIM. The memory traffic caused by error correction can be ignored compared with the overall memory traffic.

\inpnglw{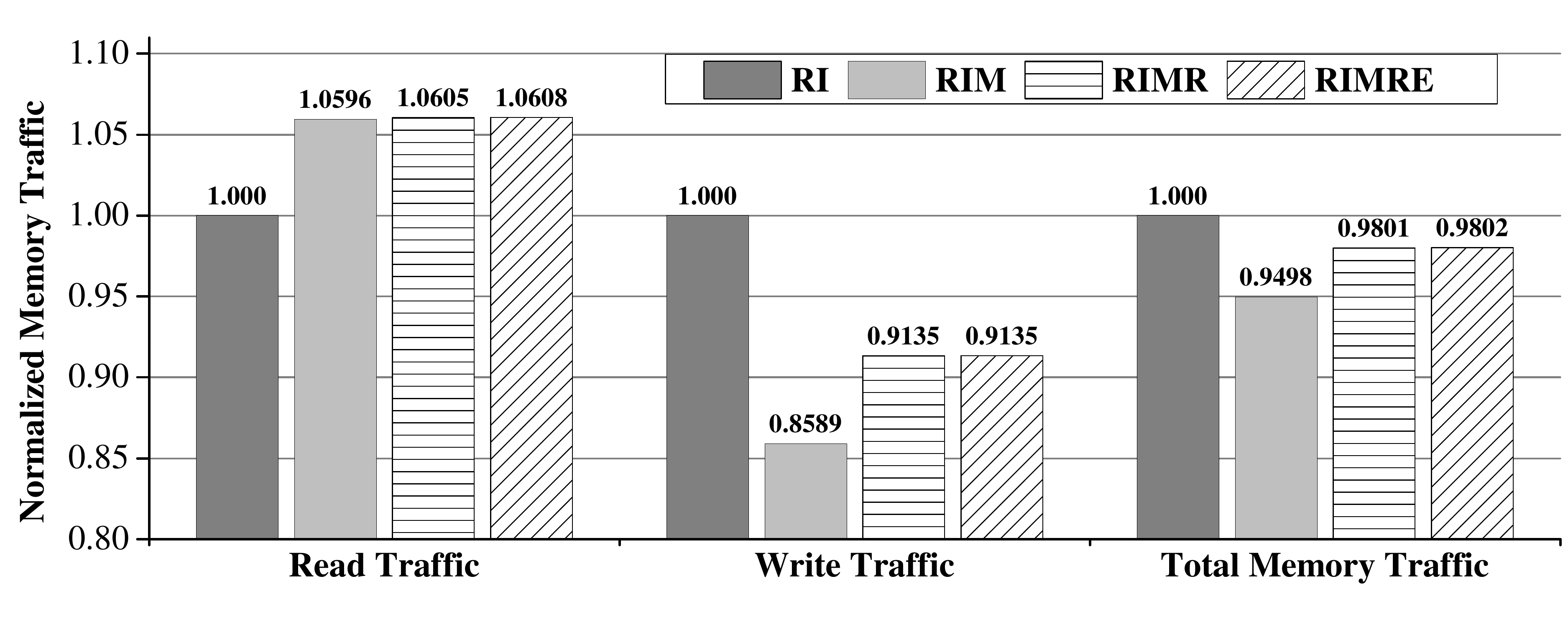}{Averaged and normalized read, write, and total memory traffic.}{EMT}{0.48}

\subsection{Impact on Energy}
In this section, we evaluate the energy efficiency of our schemes. Power-Delay Product (PDP) is the metric to evaluate the average energy consumed, but it does not reflect the speed of computing. A system with low PDP may be very slow in performing its computing. 
Energy-Delay Product (EDP) is the average energy multiplied by the time it takes to complete the computation. It is a much more preferable metric for evaluating energy efficiency than PDP.
Figure~\ref{EEDP} shows the EDP of the different configurations. RI, RIM, RIMR, RIMRE increase the EDP of the Baseline by 11.09\%, 7.03\%, 12.04\%, and 12.08\% respectively. This is decided by the number of memory requests and the total execution time.

\inpnglw{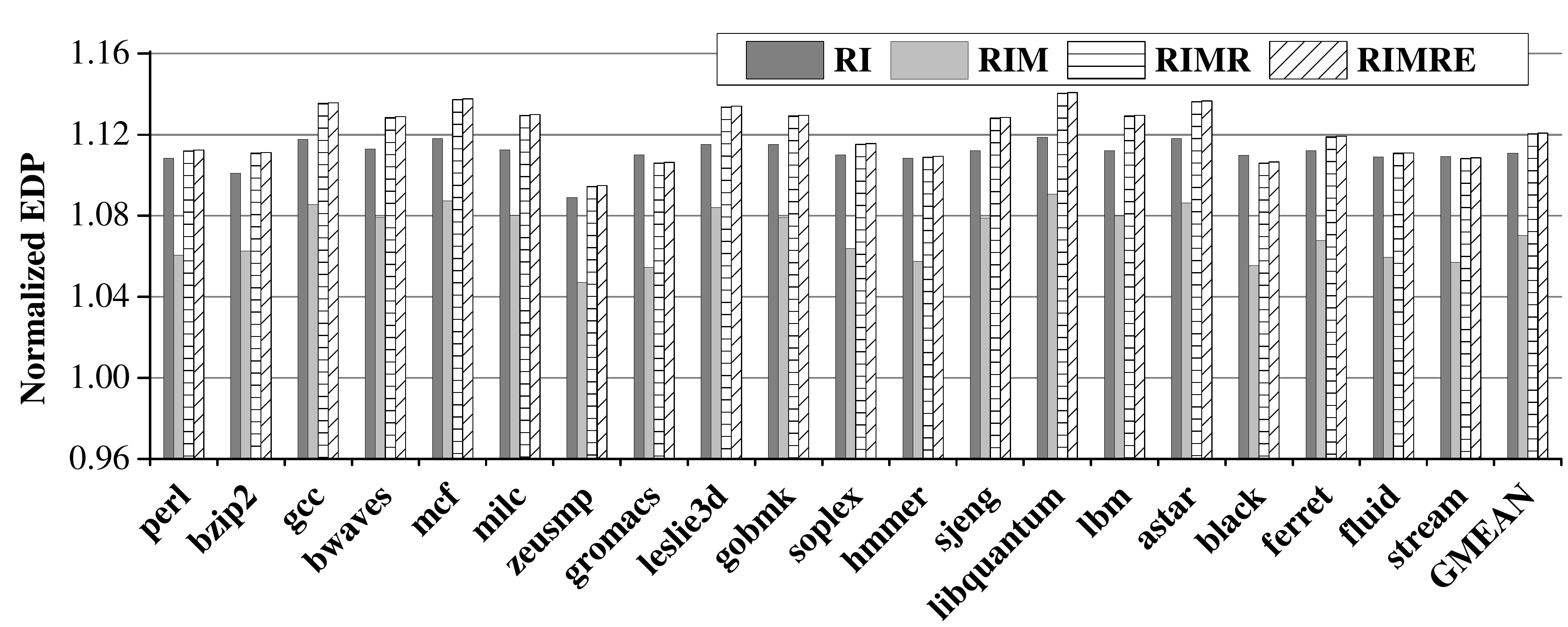}{Normalized Energy-Delay Product of RI, RIM, RIMR, and RIMRE.}{EEDP}{0.48}

\subsection{Comparison with Prior Works}
In this section, we evaluate the performance of Path ORAM, Tiny ORAM, Ring ORAM, RI, RIM, and RIMR, all normalized to the Ring ORAM (Baseline). All these ORAM schemes adopt the same hardware organization, e.g. top seven levels cached on-chip and five real slots per bucket. Figure~\ref{ECOP} shows the normalized execution time of different schemes. Compared with Ring ORAM, Path ORAM and Tiny ORAM increase the execution time by 58.91\% and 31.65\% respectively. Our proposal provides additional integrity and reliability guarantees for Ring ORAM, and only increases 7.53\% execution time which is far less than basic Path ORAM or Tiny ORAM.
\inpnglw{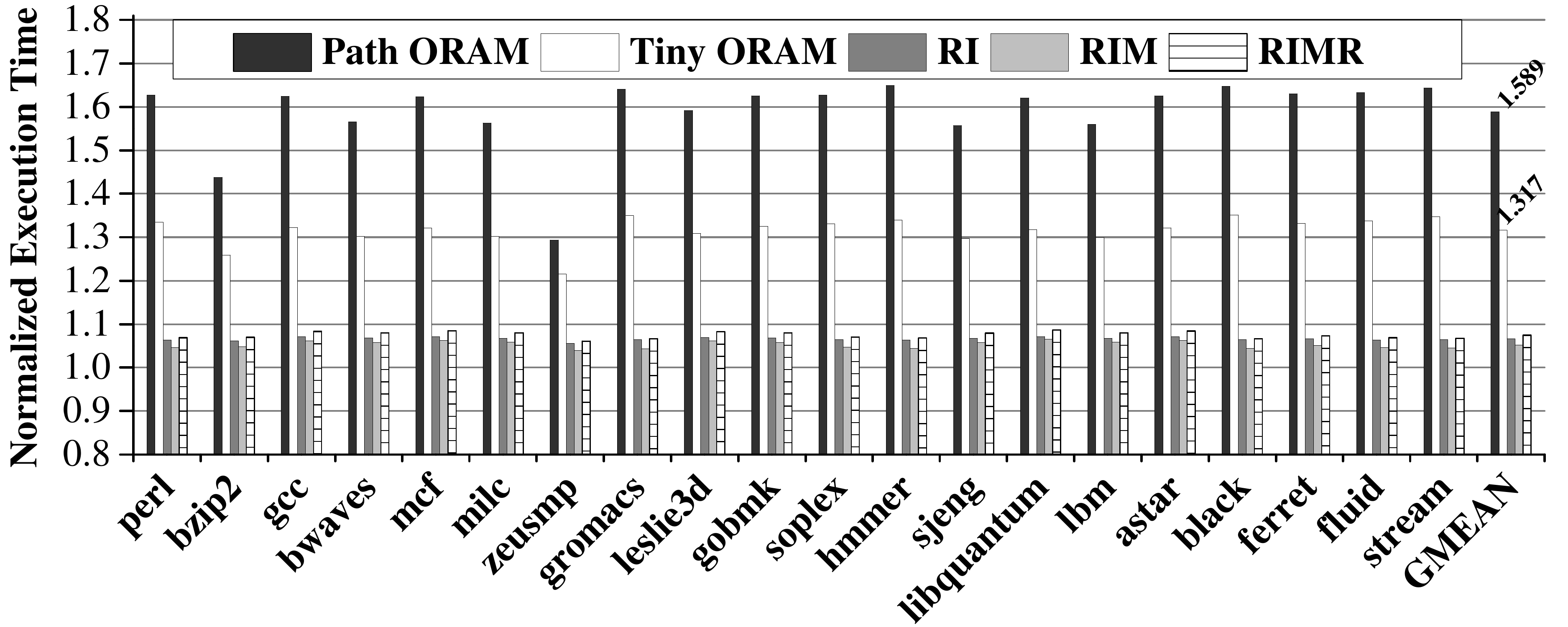}{Comparisons with prior works.}{ECOP}{0.48}

\subsection{Sensitivity to AES-GCM Unit Count} \label{sens_gcm}
In this section, we evaluate the impact of the number of AES-GCM unit on performance. 
While the decryption can be overlapped with the memory read operation, the MAC computing is started only after the data is fetched to the processor. 
Without enough AES-GCM units, the integrity verification for a fetched block would be blocked before the previous completely. 
Thus, the integrity verification would become the performance bottleneck of the computing system in such a situation. 
Figure~\ref{ESG} shows the system performance with 4, 8, and 16 AES-GCM units, and the results are normalized to the 4 AES-GCM units baseline. 
With eight AES-GCM units, the RI, RIM, and RIMR achieve 1.0049$\times$, 0.9485$\times$, and 0.9785$\times$ execution time of the Baseline. 
Compared with the four AES-GCM units setting, eight AES-GCM units reduce 6.13\%, 10.42\%, and 9.68\% execution time of RI, RIM, and RIMR respectively. 
Therefore, the main cause of performance decreases is the MAC computing congestion other than the 80-cycle MAC computing latency.
When we increase the units from 8 to 16, there was little improvement in performance. 
This means eight AES-GCM units is enough for a 2-channel and 23 levels Ring ORAM tree setting.

\inpnglw{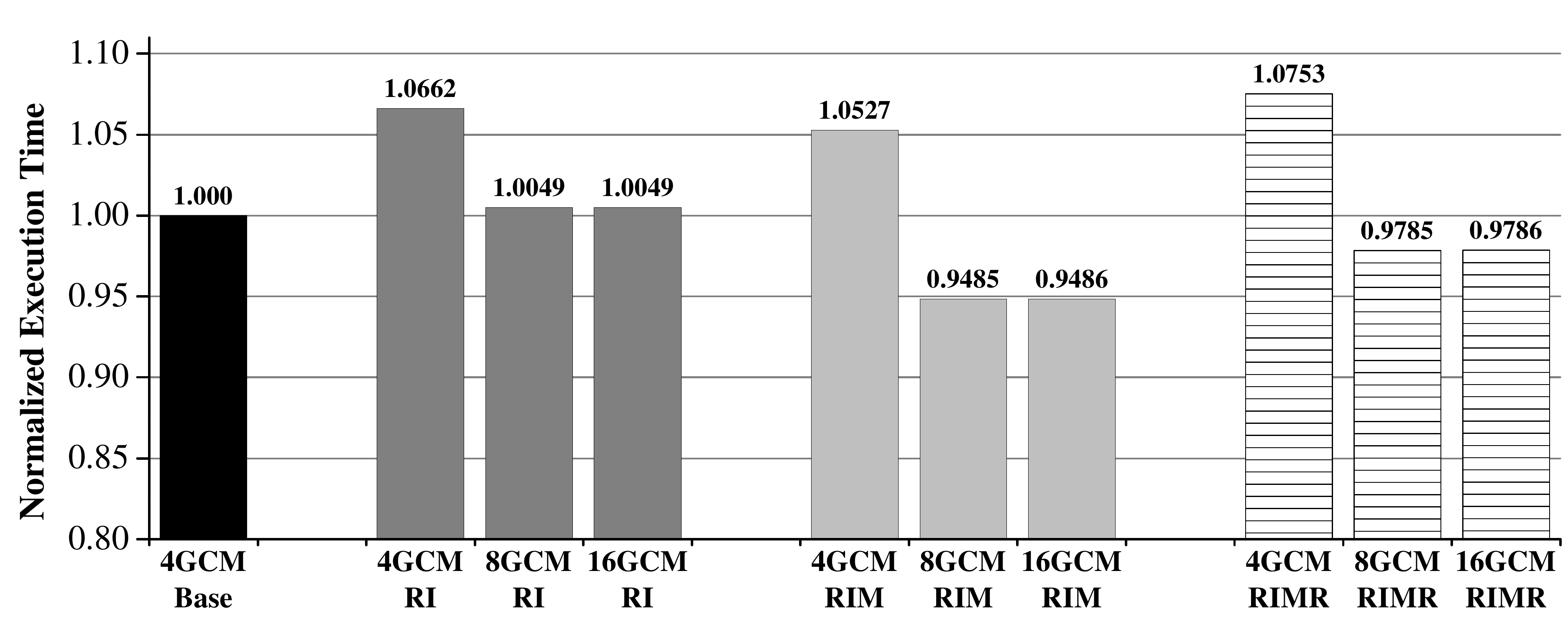}{Sensitivity to AES-GCM unit count, the results are averaged over 20 workloads and are normalized to the 2 channels 4 AES-GCM units Baseline.}{ESG}{0.48}

\subsection{Sensitivity to Channel Count}

\inpngtlw{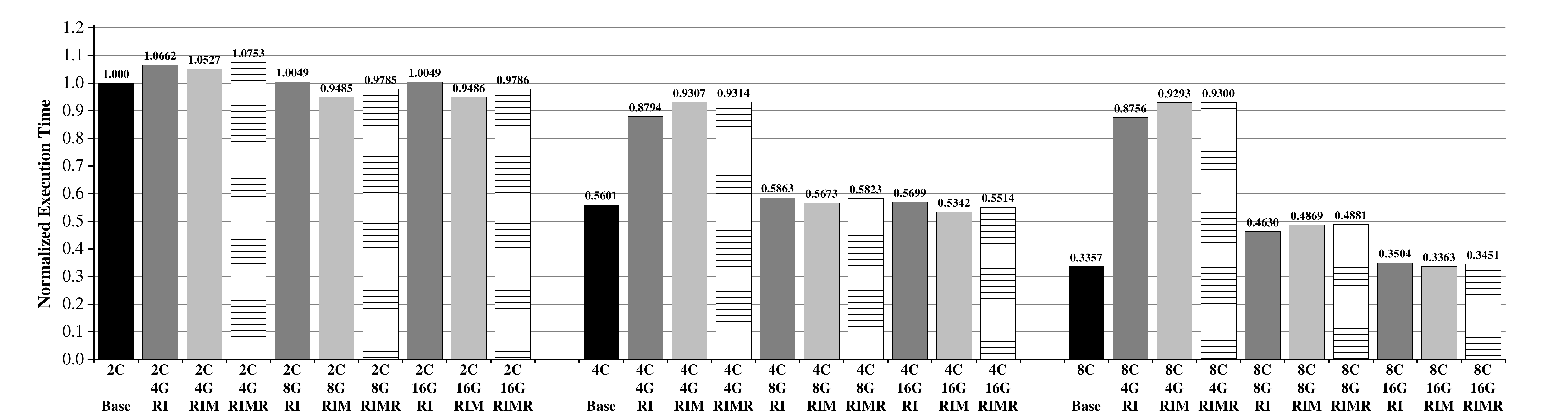}{Sensitivity to channel count and AES-GCM unit count, the results are averaged over 20 workloads and are normalized to the 2 channels Baseline.}{ESCG}{1.02}

In this section, we analyze the performance bottleneck of our schemes under the different numbers of channels and AES-GCM units.
Figure~\ref{ESCG} shows the execution time of different combinations, and all normalized to the 2 channels 4 AES-GCM units Baseline. We use the letter \textbf{C} and \textbf{G} to indicate \textbf{C}hannel and AES-\textbf{G}CM unit respectively. For example, 2C4G represents two channels and four AES-GCM units. 

From the figure, we can see that the 4C and 8C Baseline achieves 0.5601$\times$ and 0.3357$\times$ execution time of the 2C Baseline. With increasing channels, the system becomes less bandwidth bound. 
For the configurations that require integrity and reliability protection, the number of AES-GCM units is another factor that influences performance.
As discussed in the previous section, in the two-channel memory system, the four AES-GCM units setting gains an acceptable performance overhead, and more than eight units will not improve performance anymore. 

When the number of channels is increased but the AES-GCM unit count remains unchanged, the MAC computing would become the performance bottleneck of the system.
Under the 4C4G setting, 4C Baseline, RI, RIM, and RIMR achieve 0.5601$\times$, 0.8794$\times$, 0.9307$\times$ and 0.9314$\times$ execution time of the 2C Baseline. 
The RI increases 31.93\% execution time than the 4C Baseline. Obviously, MAC computing congestion is the main cause of the slowdown. 
An unexpected finding is that the use of MUST increases the execution time by 5.13\% compared with RI. 
The reason is the MUST increases the overall read traffic.
The lack of enough AES-GCM units causes more serious congestion in \emph{Read Path} when the MUST adopted. 
As the read request is on the critical path of program execution, the increased \emph{Read Path} completion time reduces the performance. 
Although the MUST decreases write traffic, it is not critical for performance in such a scenario.
The write requests are actively postponed under the most memory requests schedule policy, e.g. FCFS, FRFCFS. 
Thus, the MAC computing of write requests are less restricted by the AES-GCM resources and would not block program execution. 
When we increase the AES-GCM unit count from 4 to 8, RI, RIM, and RIMR achieve 0.5863$\times$, 0.5673$\times$, and 0.5823$\times$ execution time of the 2C Baseline. 4C8G setting can greatly alleviate the MAC computing congestion and improve performance. When we further increase the AES-GCM unit count from 8 to 16, there is only a small improvement in performance. 

In the eight-channel memory system, similar conclusions can also be obtained as in the four-channel system. 
Under 4G settings, RI/RIM/RIMR has nearly execution time in both the 4C system and 8C system. 
This result once again proves that MAC computing congestion is the main reason for the slowdown. 
When we increase the AES-GCM units from 4 to 8 and then to 16, the performance is improved responsively. 
We can see that 8G is not enough for the 8C system as the RIM spends more time to complete execution than RI. This means the congestion of MAC computing still has a greater impact on performance than the number of memory requests. 
Under the 8C16G setting, the RI, RIM, RIMR only increase the execution time of the 8C Baseline by 1.47\%, 0.06\%, and 0.94\% respectively. 
More AES-GCM units would have little performance improvement.

From the evaluation, we can get the conclusion that for a \emph{x}-channel memory system, allocate \emph{2x} AES-GCM units can gain an acceptable performance overhead. Given more AES-GCM units would not have a significant performance improvement but can reduce the execution time of RIMR to below the Baseline.

\section{Conclusion}\label{sec_concl}
In this paper, we present a Ring ORAM based security and reliability co-design - IRO. To the best of our knowledge, this is the first work to provide channel-level error resilience for the ORAM protected memory. We first present a low overhead integrity verification scheme RIT to protect the ORAM tree from replay attacks and use a MUST to decrease metadata update overhead. We then present Secure Replication to replicate real blocks to the dummy slots in the different channels to provide channel-level error resilience and use the mirrored channel technique to guarantee the reliability of the MUST. Finally, we use the ECP to repair the permanent memory cell fault, and a compact metadata design is used to reduce the storage and consulting overhead of the ECP. Through the evaluation, we can see that the performance degradation caused by our integrity and reliability protection scheme is acceptable. With the MUST optimization and enough AES-GCM units, IRO can achieve better performance than the basic Ring ORAM.

\ifCLASSOPTIONcaptionsoff
  \newpage
\fi

\bibliographystyle{IEEEtran}
\bibliography{IRO.bib} 

\end{document}